\documentclass[aps,prd,showkeys,nofootinbib,notitlepage,longbibliography,twocolumn, superscriptaddress,preprintnumbers]{revtex4-1}

\usepackage[utf8]{inputenc}
\usepackage{amsmath}
\usepackage{amsfonts}
\usepackage{amssymb}
\usepackage{color}
\usepackage{graphicx}
\usepackage{epstopdf}
\usepackage{verbatim}
\usepackage{cancel}
\usepackage{siunitx}
\usepackage{braket}
\usepackage[colorlinks,linkcolor=blue,citecolor=blue]{hyperref}

\newcommand{\dd}{\mathrm{d}}
\usepackage{fancyvrb}
\usepackage{subfig}
\numberwithin{equation}{section}

\usepackage{array,mathtools,booktabs,multirow}
\usepackage{tikz}
 
\usepackage{array}

\begin{document}
\title{Real-time dynamics of axial charge and chiral magnetic current in a non-Abelian expanding plasma}
\author{Sebastian Grieninger}
\email{sebastian.grieninger@stonybrook.edu}
\affiliation{Center for Nuclear Theory, Department of Physics and Astronomy,
Stony Brook University, Stony Brook, New York 11794–3800, USA}
\author{Sergio Morales-Tejera}
\email{sergio.moralest@uam.es}
\affiliation{Instituto de F\'isica Te\'orica UAM/CSIC, c/Nicol\'as Cabrera 13-15, Campus de Cantoblanco, 28049 Madrid, Spain}
\affiliation{Departamento de F\'isica Te\'orica, Universidad Aut{\'o}noma de Madrid, Cantoblanco, 28049 Madrid, Spain}

\preprint{IFT-UAM/CSIC-23-109}
\date{\today}

\begin{abstract}\noindent
Understanding axial charge dynamics driven by changes in Chern-Simons number densities is a key aspect in understanding the chiral magnetic effect (CME) in heavy-ion collisions. Most phenomenological simulations assume that a large amount of axial charge is produced in the initial stages and that axial charge is conserved throughout the simulation. Within an (expanding) homogeneous holographic plasma, we investigate the real-time axial charge relaxation dynamics and their impact on the chiral magnetic current. Moreover, we discuss the real-time interplay of the non-Abelian and the Abelian chiral anomaly in the presence of a strong magnetic field. In the expanding plasma, the Chern-Simons diffusion rate and thus the axial charge relaxation rate are time dependent due to the decaying magnetic field. We quantify the changes in the late time falloffs and establish a horizon formula for the chiral magnetic current. 
\end{abstract}

\maketitle

\section{Introduction}

The QCD vacuum exhibits a periodic structure, where the minima correspond to distinct Chern-Simons numbers that describe the topology of the gauge fields. An instanton or sphaleron transition~\cite{McLerran:1990de,Moore:2010jd} between such energy-degenerate vacuum sectors is
followed by a change of chirality of the chiral fermions.  The generation of chirality is a $P$- and $CP$-odd effect as was argued in~\cite{Kharzeev:1998kz,Kharzeev:2001ev,Kharzeev:2007jp,Kharzeev:2007tn}. Even though the vacuum sectors are energy degenerate they are topologically distinct. The tunneling probability from one vacuum to a topologically distinct vacuum state which is described by an instanton transition, is highly surpressed at finite temperature. However, at very high energies (for example in the initial stages of heavy ion collisions, where the quark-gluon plasma is formed~\cite{Shuryak:2004cy,Shuryak:2008eq}) we can cross the barrier(s) with a sphaleron transition to a different ground state. These transitions generate axial charge by activating the gluonic part of the axial anomaly which flips the chirality of some of the fermions and the final state is chirally imbalanced. In an external magnetic field, which aligns the spins of the chiral fermions, a change of chirality is followed by a change in the direction of momentum leading to charge separation. When there is an imbalance between the numbers of left- and right-handed fermions, this gives rise to an electric current aligned with the magnetic field direction -- the chiral magnetic effect (CME)~\cite{Fukushima:2008xe,Son:2009tf}.

The CME was first measured in condensed matter experiments \cite{Li:2014bha,li2015giant,xiong2015evidence,PhysRevX.5.031023}. In the context of heavy ion collisions, extensive experimental searches were conducted over a decade at RHIC and the LHC by ALICE and CMS culminating in the specialized RHIC experiment -- the isobar runs. This experiment utilized isobar nuclei collisions of Zr + Zr and Ru + Ru. While expecting similar collision geometries, larger magnetic fields and thus chiral magnetic currents were anticipated in Ru + Ru due to higher electric charges in Ru.

Following data collection from the RHIC isobar run, the STAR collaboration's analysis \cite{STAR:2021mii} reported no CME signal based on the predefined criteria. However, differences in collision geometry between Zr and Ru, which were not part of the predefined criteria of the isobar blind analysis, prompted a reevaluation of the isobar data through an updated analysis~\cite{Kharzeev:2022hqz,Lacey:2022plw}.
Key uncertainties influencing CME physics in the isobar run include the initial state uncertainty in collisions, affecting the heavy ion shape and proton/neutron distributions. Discrepancies in collision geometry impact the definition of centrality and charged hadron multiplicities. The charge distribution within nuclei also influences the generated magnetic field's magnitude, extent, and temporal evolution during plasma lifetime. 

The goal of this work is to incorporate Chern-Simons diffusion dynamics into real-time simulations of the chiral magnetic current at strong coupling. The resulting axial charge dynamics is usually neglected in the literature and a chirality imbalance is simply introduced by means of an axial ``chemical potential.'' 
Since chemical potentials are an equilibrium concept it is not straightforward to treat nonconserved quantities on the same footing. In particular, the divergence of the axial current is given by~\cite{Huang:2021bhj,Iatrakis:2014dka,Iatrakis:2015fma,Lin:2018nxj,Liang:2020sgr}\begin{widetext}
\begin{equation}
 \partial_m J^m_A = \epsilon^{ijkl} \left( \frac{N_f}{16\pi^2} \mathrm{tr}(G_{ij} G_{kl})+\frac{N_c
\sum_f q_f^2}{32\pi^2}F_{ij}F_{kl}
+
\frac{N_c N_f}{96\pi^2}F^5_{ij}F^5_{kl}
\right)
\end{equation}\end{widetext}
where the first term is caused by the non-Abelian anomaly chiral anomaly and the second term and third term are due to the Abelian chiral anomaly. Integrating the topological charge density $q(x)=\frac{N_f\epsilon^{ijkl}}{32\pi^2}\text{tr}(G_{ij}G_{kl})$, where $G_{ij}$ is the color field strength, we get the topological winding number $Q_W=\int\dd^4 x \,q(x)$, which characterizes the different vacua.
The topological charge density is related to the axial charge relaxation rate by
\begin{equation}\label{eq:q5relax}
    \frac{\dd n_5}{\dd t}=-2q=-\frac{2\Gamma_\text{CS}}{\chi_5\,T}n_5=-\frac{n_5}{\tau_\text{sph}},
\end{equation}
where the factor $\tau_\text{sph}$ is related to the relaxation time of topological charge fluctuations~\cite{Huang:2021bhj} and $\chi_5$ is the axial susceptibility. The last two equalities in eq. \eqref{eq:q5relax} are linear response expressions which are valid at late times close to equilibrium. The
axial charge relaxation time is related to the Chern-Simons diffusion rate $\Gamma_\text{CS}$ and given by
\begin{equation}\label{eq:CSdiff}
    \tau_\text{sph}=\frac{\chi_5 T}{2\,\Gamma_\text{CS}}.
\end{equation}
Equation~\eqref{eq:CSdiff} relates the Chern-Simons diffusion rate $\Gamma_\text{CS}$ to the axial susceptibility $\chi_5$ and sphaleron rate $\tau_\text{sph}$ (which is related to the axial charge relaxation rate). The Chern-Simons diffusion and axial charge relaxation rate are both accessible in the homogeneous limit, i.e. at zero wave vector. Moreover, since the $U(1)_A$ symmetry is explicitly broken by the dynamical non-Abelian gauge fields, the chiral magnetic wave~\cite{Kharzeev:2010gd} is gapped in the homogeneous limit and axial charge relaxation is incorporated in the homogeneous dynamics.\footnote{Instead of the pair of propagating sound modes, we observe one diffusive hydrodynamic mode (associated with the unbroken $U(1)_V$ symmetry) and a gapped pseudo-diffusive mode (associated with the explicitly broken $U(1)_A$) which governs the relaxation of axial charge.}

Important insight into the topological dynamics were achieved in terms of classical statistical simulations \cite{Schlichting:2022fjc,Mace:2016svc,Muller:2016jod,Mace:2016shq} featuring the non-Abelian anomaly. On the holographic side, out-of-equilibrium simulation of the CME (\cite{Yee:2009vw,Gynther:2010ed,Landsteiner:2013aba}) in an infinite, static plasma were first performed in \cite{Lin:2013sga,Ammon:2016fru,Grieninger:2016xue} (probe limit) and \cite{Ghosh:2021naw,Grieninger:2021zik} (including backreaction). This was recently extended to an expanding plasma in \cite{Cartwright:2021maz}.\footnote{Note that there is also an extensive list of holographic works on the axial CME where a $U(1)_A$ symmetry instead of the $U(1)_A\times U(1)_V$ symmetry is considered~ \cite{Haack:2018ztx,Ammon:2016szz,Ammon:2017ded,Ammon:2020rvg,Cartwright:2020qov}.} In holography, the axial charge relaxation and Chern-Simons diffusion rate were discussed in \cite{Grieninger:2023wuq,Gursoy:2012bt,Iatrakis:2014dka,Jimenez-Alba:2014iia,Gursoy:2014ela,Drwenski:2015sha,Iatrakis:2015fma,Jimenez-Alba:2015awa,Bigazzi:2018ulg,Gallegos:2018ozs,Craps:2012hd,Bu:2014cca,Jahnke:2014sla,Son:2002sd,Hou:2017szz}. In this work, we will generalize the results to include both the Abelian chiral anomaly and the non-Abelian anomaly as well as strong (external) Abelian magnetic fields and dynamical non-Abelian gauge fields. The appropriate holographic model, to incorporate the dynamical gauge field contributions due to the dynamical gluons is the so-called St\"uckelberg model where the non-Abelian gluon dynamics is coupled to the axial gauge field via a $\theta$ term -- rendering the axial gauge field massive \cite{Klebanov:2002gr,Maldacena:2000yy,Anastasopoulos:2006cz,Jimenez-Alba:2014iia}. The St\"uckelberg (pseudo) scalar is the holographic analog of the $\theta$ term in QCD. In this work, we use the minimal bottom-up model developed in~\cite{Jimenez-Alba:2014iia},\footnote{For other works using the St\"uckelberg model see for example \cite{Megias:2019djo,Rai:2023nxe,Baggioli:2023tlc,Grieninger:2023wuq}.} where the axial current is nonconserved and acquires an anomalous dimension.

In heavy-ion collisions, the plasma starts expanding rapidly after the collision. In order to make connection to phenomenology, we extend our results for the static plasma in section \ref{sec:static} to an expanding plasma \ref{sec:Expanding}. In the expanding plasma, the energy density, axial charge density and magnetic field decay due to the dilution. In particular, some models in magnetohydrodynamics indicate that -- at late times -- the magnetic field decays inversely proportional to the proper time $B\sim \tau^{-1}$~\cite{Deng:2012pc,Yan:2021zjc,Pu:2016ayh,Roy:2015kma,Siddique:2019gqh}. If we assume that the axial charge is conserved, we expect the axial charge density in our holographic plasma to decay similarly ($n_5\sim\tau^{-1}$). 
This also impacts the CME that is within linearized hydrodynamics proportional to the axial charge density divided by the susceptibility and the magnetic field B, suggesting a decay proportional to $\sim \tau^{-4/3}$. These expressions are only valid if the axial charge is conserved. As we will show in section \ref{sec:expandingtwo} the decay will be accelerated due to the explicit breaking of the $U(1)_A$. The crucial question becomes whether the chiral magnetic current has enough time to build up sufficiently in magnitude during the plasma phase to manifest as measurable signal. Our model aims to address this comprehensively.

The outline of the paper is the following. In section \ref{sec:model}, we review the holographic model used for the simulations. We then study an infinite, static plasma where we investigate the influence of Abelian and non-Abelian anomalies and the magnetic field on the axial charge relaxation time in section \ref{sec:qnm}. In section \ref{sec:staticplasmadyn}, we incorporate axial charge generation and measure the real-time CME response to it. Moreover, we perform a parameter scan. Finally, in section \ref{sec:expandingone} and \ref{sec:expandingtwo}, we study the dynamics in an expanding plasma as it is phenomenologically relevant for heavy-ion collisions. The conclusions are outlined in section \ref{sec:conclusions}.

\section{Holographic model}
\label{sec:model}

We study the real-time dynamics of the chiral magnetic effect including axial charge relaxation due to Chern-Simons diffusion dynamics. We consider the minimal model established in~\cite{Jimenez-Alba:2014iia} which includes all the basic ingredients. In particular, it contains an axial gauge field $A$ and a vector gauge field $V$ dual to the Abelian axial and vector currents. The effect of the $U(1)^3_A$ and $U(1)_A\times U(1)_V^2$ anomalies is reproduced via a Chern-Simons term with appropriate coefficients. The gluonic contribution to the chiral anomaly are introduced by making the axial gauge field massive via the St\"uckelberg mechanism. Thus, the holographic action is given by

\begin{widetext}
\begin{equation*}\label{eq:actionSt}
     \!S\!=\!\dfrac{1}{2 \kappa_5^2}\! \int\limits_{\mathcal{M}}\!\! \dd^5x \sqrt{-g}\left[R+\dfrac{12}{L^2}\!-\!\dfrac{1}{4}(F^2\!+\!F_{(5)}^2)\! -\! \dfrac{m_s^2}{2}(A_{\mu}-\partial_{\mu}\theta)^2\!%(A^{\mu}-\partial^{\mu}\theta)%\right.\\&+\left. 
    +\! \dfrac{\alpha}{3} \epsilon^{\mu\nu\rho\sigma\tau}\! (A_\mu\!-\!\partial_{\mu}\theta)\left(3 F_{\nu\rho}F_{\sigma\tau} + F_{\nu\rho}^{(5)}F_{\sigma\tau}^{(5)}\right)\right] + S_\text{GHY}+S_\text{ct},
\end{equation*}\end{widetext}

\noindent
where $S_\text{GHY}$ is the Gibbons-Hawking-York boundary term to make the variational problem well defined, $L$ is the anti-de Sitter (AdS) radius, $\kappa_5^2$ is the Newton constant, $\alpha$ the Chern-Simons coupling and $m_s$ the mass of the gauge field. The Levi-Civita tensor is defined as $\epsilon^{\mu \nu \rho \sigma \tau}=\epsilon(\mu \nu \rho \sigma \tau)/\sqrt{-g}\,$. The St\"uckelberg field is denoted as $\theta$ whereas the field strengths are defined as $F=dV$ and $F_{(5)}=dA\,$. The coupling $\theta(x^{\mu})$ couples the operator $\text{Tr}\{G\wedge \tilde{G}\}$ (i.e. to gluons living on a D3 brane) thus playing the role of the $\theta$ angle (see section IV of \cite{Gursoy:2014ela}). Note that the gluon field strength does not appear explicitly in our holographic model but is mediated through the coupling $\theta$. The axial gauge field couples to the axion through the mass term and hence the dual axial current is nonconserved due to the non-Abelian anomaly. Moreover, the axial gauge field couples to the vector gauge field through the Chern-Simons term, which accounts for the Abelian anomaly.
The holographic renormalization of this model was done in \cite{Jimenez-Alba:2014iia} and the counterterm action (with restriction $\Delta<\frac{1}{3}$\footnote{\label{ft:1} This range for $\Delta$ minimizes the number of counterterms. For $\Delta\to1$, the number of counterterms diverges, so that the model is no longer renormalizable. See \cite{Jimenez-Alba:2014iia} for more details.}) reads

\begin{align}
    S_\text{ct} = \int_{\partial \mathcal{M}}& \dd^4 x \sqrt{|\gamma|}\left( \frac{\Delta}{2}(A^i-\partial^i \theta)(A_i - \partial_i \theta)  \right.\nonumber\\&- \frac{1}{4(\Delta+2)}\partial_i(A^i-\partial^i \theta)\partial_j(A^j-\partial^j \theta)\nonumber\\&
    \left. + \frac{1}{8 \Delta} F^{5\,ij}F^5_{ij} - \frac{1}{4} F_{ij}F^{ij}\log(u) \right)\,,
\end{align}

\noindent
where $\Delta \equiv -1+\sqrt{1+m_s^2}$. $\partial \mathcal{M}$ is the boundary of spacetime $\mathcal{M}$, $\gamma$ is the determinant of the induced metric in $\partial \mathcal{M}$. Latin indices refer to the boundary coordinates.  

The equations of motion derived from \eqref{eq:actionSt} read

\begin{equation}
\begin{split}
\label{eq:eoms}
    &m_s^2\nabla_{\mu}(A^\mu-\partial^\mu\theta)=0\\
    &  \nabla_\nu F^{\nu\mu}+2\alpha \epsilon^{\mu\nu\rho\sigma\tau} F_{\nu\rho}F^{(5)}_{\sigma\tau}=0\,,\\
    & \nabla_\nu F_{(5)}^{\nu\mu}\!-\!m_s^2(A^\mu -\partial^\mu\theta)\!+\!\alpha  \epsilon^{\mu\nu\rho\sigma\tau}\!\! \left( F_{\nu\rho}F_{\sigma\tau}\!+\!F_{\nu\rho}^{(5)}F_{\sigma\tau}^{(5)} \right)\!=\!0, \\
    &  G_{\mu\nu}-\frac{6}{L^2} g_{\mu\nu}\!-\!\frac{1}{2} F_{\mu\rho}F_{\nu}^{\ \rho }
 \!+\!\frac{1}{8} (F^2 + F_{(5)}^{2})g_{\mu\nu}\! -\frac{1}{2} F^{(5)}_{\mu\rho}F_{\nu}^{(5) \rho } \\&-\dfrac{m_s^2}{2}(A_\mu - \partial_{\mu}\theta)(A_\nu - \partial_{\nu}\theta)+\dfrac{m_s^2}{4} (A_\alpha-\partial_{\alpha}\theta)^2g_{\mu\nu}=0\,.
\end{split}
\end{equation}

Each expectation value of the dual field theory shall be extracted following the holographic prescription, i.e. varying the renormalized on-shell action with respect to the boundary value of the dual field. The one-point functions are given by

\begin{equation}\label{eq:1pt1}
    2\kappa_5^2\langle J^i_V\rangle\! =\!\! \left. n_{\rho} \sqrt{|\gamma|} \left( F^{i\rho}\! +\! 4\alpha \epsilon^{\rho ijkl}(A_j - \partial_j \theta)F_{kl} \right)\! +\! \dots\right|_{\partial \mathcal{M}}
\end{equation}
\begin{equation}\label{eq:1pt2}
    2\kappa_5^2\langle J^i_A\rangle =\! f(x^{\rho}) \left.  \sqrt{|\gamma|} \left( F_5^{i \rho}  n_{\rho} + \Delta (A^i-\partial^i \theta) \right) + \dots\right|_{\partial \mathcal{M}}
\end{equation}

\noindent
where the dots indicate terms that vanish when evaluated at the boundary. $n_\rho$ is the (outward pointing) normal vector to $\partial \mathcal{M}$. It is assumed that the boundary is defined by $x^{\rho}=constant\,$, where $x^{\rho}$ is the radial coordinate. The prefactor $f(x^{\rho})$ is the variation of the $i-th$ component of the axial gauge field with respect to its non-normalizable mode. That means, if $A$ behaves as $A \sim A_{NN} u^{-\Delta} + \text{higher orders in }u$ near the boundary,  then we have $f(x^{\rho})= u^{-\Delta}\,$ plus higher orders which do not contribute to the expectation value.

\subsection{Static plasma}\label{sec:static}
Let us first discuss the setup in a static plasma in detail which will allow us to skip most of the details in the case of an expanding plasma. We switch on a constant and homogeneous magnetic field $B$. In the static case, the system is symmetric under shifts of the three-dimensional spatial coordinates, which we denote by $(x,y,z)$, and hence the metric functions will only depend on the radial and temporal coordinates $u$ and $v$ respectively.  The magnetic field breaks the $SO(3)$ rotational invariance down to $SO(2)$ causing anisotropy even at equilibrium. Taking $B$ to point in the $z$-th direction preserves rotational invariance in the $x$-$y$ plane. In infalling Eddington-Finkelstein coordinates our ansatz is~\cite{Chesler:2013lia,Fuini:2015hba}
\begin{align}\label{eq:II.6}
    \dd s^2=&-f(v,u) \dd v^2 -\frac{2L^2}{ u^2} \dd u\, \dd v \\&+ \Sigma(v,u)^2\left(e^{\xi(v,u)}(\dd x^2+\dd y^2)+e^{-2\xi(v,u)}\dd z^2\right)\,,\nonumber
\end{align}

\noindent
where $f$, $\Sigma$ and $\xi$ are undetermined functions of $u$ and $v$. The boundary is located at $u=0\,$. We further demand that the metric asymptotes to $AdS_5$: 

\begin{equation}
    \lim_{u\to 0}f=\frac{L^2}{u^2}\,, \hspace{0.3cm} \lim_{u\to0}\Sigma = \frac{L}{u}\,, \hspace{0.3cm} \lim_{u\to0}\xi = 0\,.
\end{equation}

The chiral magnetic current builds up in the presence of a magnetic field and chiral imbalance. The chiral imbalance is introduced switching on the temporal component of the axial gauge field $A$\footnote{We work in the radial gauge, namely the radial components of both gauge fields are set to zero.}. The chiral magnetic current is parallel to the magnetic field, so it will also be aligned in the $z$-direction and a consistent solution requires switching on the $z$ component of the vector gauge field $V$. Finally, the St\"uckelberg field $\theta$ needs also be switched on. To sum up, we write
\begin{equation}\label{eq:II.8}
\begin{split}
   & A = -A_t(v,u)\dd v\,,\ \ \theta=\theta(v,u)\,,\\
   & V = \frac{B}{2}(x\, \dd y-y\,\dd x)+V_z(v,u)\, \dd z\,.\\
 %  & \,.
\end{split}
\end{equation}
We do not source the currents in the dual field theory, which amounts to setting the leading modes of the gauge fields in the near boundary expansion to zero. Then, the asymptotic solution to the equations of motion reads
\begin{equation}\label{eq:II.9}
\begin{split}
    &A\simeq \dot{\theta}_0(v) + u^\Delta \left(q_5(v)u^2 + \frac{2+\Delta}{3+\Delta} \Dot{q}_5(v)u^3 + \mathcal{O}(u^4)\right) \\
    &V\simeq u^2 V_2(v)+\mathcal{O}(u^3)\\
    &\theta \simeq \theta_0(v) + \mathcal{O}(u^{3+\Delta})+ \mathcal{O}(u^{5})\\
    &\xi \simeq u^4 \left(\xi_4 -\frac{B^2}{12}\log u \right) + \mathcal{O}(u^5)\\
    &\Sigma\simeq \frac{1}{u} + \lambda(v) + \mathcal{O}(u^5)\\
    &f \simeq \left(\frac{1}{u}+\lambda(v)\right)^2 - 2 \dot{\lambda}(v) + u^2(f_2+\frac{B^2}{6}\log u) + \mathcal{O}(u^3)
\end{split}
\end{equation}
\noindent
Powers of $u^{n \Delta}$ for integer $n$ appear at higher order in the expansion for all fields due to mixing. The coefficients $n_5\,,V_2$ and $f_2$ are related to the expectation values of operators in the dual field theory. The expectation value $\theta_4$ of the axion field is read off from the order $u^4$. It is proportional to the time derivative of the source of $V$, which is zero in our ansatz. The source for the axion $\theta_0$ is a remnant of gauge invariance and may be set to zero without loss of generality.

Substituting the asymptotic expansions into equations \eqref{eq:1pt1} and \eqref{eq:1pt2} gives 
\begin{align}\label{eq:II.10}
    J_{CME} &\equiv 2\kappa_5^2 \left<J^z\right> = 2 V_2(v), \\  n_5&\equiv 2\kappa_5^2 \left<J^t_5\right>  = 2(1+\Delta) q_5(v),\label{eq:II.10b}
\end{align}
\noindent
which correspond to the chiral magnetic current and the axial charge, respectively. Note that the scaling dimension of the axial current in the dual theory is $3+\Delta$. As a consequence, having $\Delta>1$ renders the dual QFT is nonrenormalizable. This fact goes hand in hand with footnote \ref{ft:1}.

In general $\Delta$ will be a noninteger causing noninteger powers of $u$ in the near boundary expansion. This, along with the presence of six undetermined fields makes the numerical problem harder to deal with. We can simplify the problem considerably by taking into account the conditions relevant for the quark-gluon plasma phenomenology. In particular, high estimations of axial charge $n_5$ reach only up to $n_5/s\sim 0.1$ (see for instance  \cite{Shi:2018izg,Huang:2017tsq}), where $s$ is the entropy density of the plasma. With this in mind, we take the small axial charge limit: $n_5 \to \epsilon n_5$.  In our ansatz this amounts to 
\begin{equation}
\begin{split}\label{eq:ansatzmatter}
   & A = - \epsilon A_t(v,u)\dd v,\  \theta=\epsilon\, \theta(v,u)\,,\\
   & V = \frac{B}{2}(x\,\dd y-y\, \dd x)+\epsilon V_z(v,u)\, \dd z\,,
\end{split}
\end{equation}
\noindent
and then solve the equations at zeroth and first order in $\epsilon$. In appendix \ref{app:smallcharge}, we show that the error caused by this approximation is small by comparing the small charge evolution to the full result (at $\Delta=0$). In the nonexpanding case, the zeroth order equations 
\begin{equation}\label{eq:STbackground}
\begin{split}
    & \xi'^2+\frac{2 \left(u^2 \Sigma'\right)'}{u^2 \Sigma}=0\,,\\
    &\frac{3  \Sigma'}{\Sigma} f'+\left(\frac{6 
   \Sigma'^2}{\Sigma^2}-\frac{3}{2}  \xi'^2\right)f+\frac{1}{2}\frac{B^2 e^{-2 \xi}}{ u^4 \Sigma^4}-\frac{12}{u^4}=0\,,\\
   & f \xi''+\left(\frac{ f'}{ f}  +\frac{3  \Sigma' }{
   \Sigma}\!+\!\frac{2  }{u}\right)f\xi'+\frac{1}{3}\frac{B^2 e^{-2 \xi}}{ u^4 \Sigma^4}=0\,,\\
   &f''\!+\!\left(\frac{4  \Sigma'}{\Sigma}\!+\!\frac{2
   }{u}\!\right)\!f'\!+\!\left(\frac{2  \Sigma'^2}{\Sigma^2}\!-\!\frac{1}{2} 
   \xi'^2\!\!\right)\!f-\!\frac{B^2 e^{-2 \xi}}{6 u^4 \Sigma^4}\!-\!\frac{12}{u^4}=0,
\end{split}
\end{equation}
\noindent
are solved by a static background configuration for the metric. We denoted derivatives with respect to the radial coordinate $u$ with a prime. The scale $L$ has been set to $1$. To find the background configuration, we have to solve for the metric fields $(f,\xi,\Sigma)$ with appropriate boundary conditions. In particular, we demand AdS$_5$ asymptotics and the presence of a regular horizon which we choose to be at $u_h=1$. Since the background is static we may unambiguously compute the temperature of the black hole, which matches that of the dual field theory.
\begin{equation}
    T = \frac{1}{2\pi}\left. \left(-\frac{u^2}{2}\partial_u f \right)\right|_{u_h}\,.
\end{equation}
The four equations are not independent, the last one is implied by the first three equations and their radial derivatives. Note that the background is static and takes into account the presence of the magnetic field but is uncharged. Einstein's equations at first order in $\epsilon$ are trivially satisfied, corrections appear at order $\epsilon^2$ and are neglected at linear order. The matter equations of motion however are nontrivial at first order in $\epsilon$:
\begin{equation}\label{eq:eomstatic}
\begin{split}
    &A_t''+\left(\frac{3
   \Sigma'}{\Sigma}+\frac{2}{u}\right) A_t'-\frac{\Delta\left(\Delta
   +2  \right) }{u^2}\theta'+\frac{ 8 \alpha  B}{u^2 \Sigma^3}V_z'=0\,,\\
   &\Dot{A}_t'+\frac{\Delta\left(\Delta+2  \right)
   }{u^2}\left(A_t-\Dot{\theta}+u^2 f \theta'\right)+\frac{8 \alpha  B }{u^2 \Sigma^3}\Dot{V}_z=0\,,\\
   &\Dot{V}_z'+\Dot{V}_z \left(\frac{\Sigma'}{2
   \Sigma}+\xi'\right)-\frac{1}{2}u^2fV_z' \left( \frac{f'}{f} +\frac{ \Sigma'}{ \Sigma}+
   2\xi'+\frac{2}{u} \right)\nonumber\\&-\frac{1}{2} u^2 f V_z''-\frac{4 \alpha  B }{e^{2 \xi}\Sigma }A_t'=0\,,\\
   &\Dot{\theta}'+\Dot{\theta}\frac{3 \Sigma' }{2
   \Sigma}-\frac{1}{2} u^2 f \theta''-\frac{1}{2}u^2 f\theta' \left( \frac{f'}{f} +\frac{3  \Sigma'}{
   \Sigma}+\frac{2}{u}\right)\nonumber\\&+\frac{1}{2} A_t'+\frac{3
   \Sigma' }{2 \Sigma}A_t=0\,.
\end{split}
\end{equation}
\noindent
We denote time derivatives with an overdot. The previous equations are not independent, in particular the time derivative of the first one is implied by the other three equations and their radial derivatives. Thus, the first equation may be regarded as a constraint on the initial data. The metric fields $(f,\Sigma,\xi)$ are those obtained solving the Einstein's equations at zero order in $\epsilon$. Recall that these fields contain the information regarding the magnetic field. We linearize the matter fields $(A_t,V_z,\theta)$ to first order in $\epsilon$, which means that we may rescale all fields by the same amount and we still get a valid solution. These rescalings will later play an important role when we discuss the initial state.

A valid initial state is specified by giving a profile to two of the three matter fields. In particular, we shall start with a trivial profile for $V_z$: $V_z(0,u)=0\,$; and with $A_t(0,u) = n_5(0) u^{2+\Delta} + \frac{2+\Delta}{3+\Delta} \Dot{q}_5(0)u^{3+\Delta}$. The initial profile for $\theta$ is obtained solving the constraint equation, i.e. the equation without time derivatives. Physically this corresponds to a nonequilibrium state in which there is no chiral magnetic current and there may be some amount of axial charge whose initial time evolution is captured by $\Dot{q}_5(0)$.

\subsection{Expanding plasma}\label{sec:Expanding}
In this section we detail the ansatz for a boost invariant expanding plasma which is phenomenologically more relevant than the case of the static plasma.
In order to be consistent with the notation of the last section, we chose the magnetic field to point in the $z$ direction and the plasma is expanding along the $\hat\eta$ direction. At the conformal boundary, we demand that the boundary metric is of the form~\cite{Chesler:2009cy,Critelli:2018osu,Rougemont:2021qyk,Cartwright:2021maz,Rougemont:2022piu}
\begin{equation}\label{eq:bdym}
 \lim\limits_{u\to0} \frac{L^2}{u^2} \, \dd s^2=-\dd \tau^2+\tau^2\dd\hat\eta^2+\dd y^2+\dd z^2
\end{equation}
which can be achieved by making the following ansatz
\begin{align}
&\!\!   \dd s^2\!=-f(\tau,u) \dd v^2 -\frac{2L^2}{ u^2} \dd u\, \dd \tau \\&
\!\!\!+\! \Sigma(\tau,u)^2\left(\!e^{-\xi_1(\tau,u)-\xi_2(\tau,u)}\dd \xi^2\!+\!e^{\xi_1(\tau,u)}\dd y^2\!+\!e^{\xi_2(\tau,u)}\dd z^2\!\right)\!.\nonumber
\end{align}
To recover the metric \eqref{eq:bdym} at the boundary, we impose  (at $u= 0$)
\begin{align}
    f=\frac{L^2}{u^2},\,\Sigma=\frac{\tau^{1/3}}{L^{4/3}\,u},\,\xi_1=\xi_2=-\frac{2}{3}\log\left(\frac{\tau}{L}\right).
\end{align}
The ansatz for the matter fields is the same as in eq.~\eqref{eq:ansatzmatter}. Note that due to the expansion of the plasma the magnetic field and axial charge (at $\Delta=0$) decay with $1/\tau$ due to dilution. As we will see, in this work the relaxation of the axial charge is modified at finite $\Delta$. Similar to the static case, we solve the equations of motion to first order in $\epsilon$. The main difference is that the background is time dependent due to the expansion of the plasma and the energy density (and hence temperature), longitudinal and transverse pressure, magnetic field and axial charge decay with time. 

In order to relate our parameters to three-flavor QCD, we follow our matching procedure outlined in~\cite{Ghosh:2021naw,Grieninger:2021rxd,Grieninger:2021zik}. In the infinite plasma, the entropy density is proportional to the area of the black hole horizon $A_\text{BH}$
\begin{equation}
    s=\frac{A_\text{BH}}{4G_N\,\text{Vol}(\mathbb R^3)}=\frac{2\pi}{\kappa_5^2}\left(\frac{L}{u_h}\right)^3=\frac{2\pi^4\,L^3}{\kappa_5^2}T^3,
\end{equation}
where $u_h$ is the black hole horizon. Similarly, the expanding background asymptotes to the Bjorken expanding plasma at very late (proper) time whose gravitational dual was introduced in~\cite{Janik:2005zt,Janik:2006gp}. The dual geometry may be viewed as a black hole whose horizon is moving away from a boundary observer. As discussed in~\cite{Grieninger:2022yps,Kim:2007ut}, the asymptotic Bjorken geometry may be mapped onto a static black hole with metric
\begin{align}
    \mathrm{d}s^2=&\frac{\pi^2 T_0^2\,L^2}{\rho}\left(-f(\rho)\frac{\tau^2_0}{t_0^2}\mathrm{d}t^2+\frac 49\,t^2\frac{\tau^2_0}{t_0^2}\mathrm{d}\eta^2+\frac 32\frac{t_0}{t}\mathrm{d}x_\perp^2\right)\nonumber\\&+\frac{L^2}{4\,f(\rho)}\frac{\mathrm{d}\rho^2}{\rho^2},\label{eq:staticBjorken}
\end{align}
where $f(\rho)=1-\rho^2, \rho$ is a mix of the original radial coordinate and proper time, $(t/t_0)=3/2(\tau/\tau_0)^{2/3}$ and the subscript zero refers to the initial values of the respective quantities. The entropy associated with~\eqref{eq:staticBjorken} reads
%\begin{equation}
  $ s=2\pi^4\, L^3\,  T_0^3/\kappa_5^2$. 
%\end{equation}
Note that even though in the Bjorken expanding plasma the temperature is proper time dependent, the entropy density per invariant volume is constant in time \cite{Janik:2005zt,Janik:2006gp,Heller:2011ju,Heller:2012je}.
On the field theory side, recall that the Stefan-Boltzmann value of the entropy density $s_\text{SB}$ is 
\begin{equation}
    s_\text{SB}=4\,\left (\nu_b+\frac{7}{4}\nu_f\right)\,\frac{\pi^2T^3}{90},
\end{equation}
where $\nu_b=2(N_c^2-1)$ and $\nu_f=2\,N_c\,N_f$ with $N_c=3=N_f$. Moreover, the axial anomaly of three-flavor QCD is given by
\begin{equation}
    \mathcal A_\text{QCD}=\frac{N_c}{32\pi^2}\,2\,\left(\frac 49+\frac 19+\frac 19\right)=\frac{1}{8\pi^2}.
\end{equation}
The Stefan-Boltzmann value is only reached at asymptotically high temperatures. Thus, we take the relative factor of 3/4 which arises in the match of the black hole entropy of gravitational
models to $\mathcal N = 4$ super-Yang-Mills plasma (at infinite ’t-Hooft
coupling)~\cite{Gubser:1996de}. Taking $3s_\text{SB}/4 = s_\text{BH}$, yields the matching conditions for Newton's constant and the Chern-Simons coupling
\begin{equation}\label{eq:newton}
   \frac{\alpha}{2\kappa_5^2}=\frac{1}{8\pi^2}, \qquad \kappa_5^2=\frac{24\pi^2 }{19}.
\end{equation}
Note that the choice of $\alpha$ depends on the value of Newton's constant and we chose the parameters so that they resemble the physics of 3-flavor QCD. Our choice of $\alpha=6/19$ does not correspond to the same value for the strength of the anomaly as in~\cite{Cartwright:2021maz} due to different choices for $\kappa_5^2$.

\section{Numerical results}

In this section we start our discussion, with the infinite, static plasma by solving the equations \ref{eq:eoms} numerically. First, we discuss the so-called quasinormal modes which capture the late time behavior of the system close to equilibrium. We continue by discussing the explicit temporal evolution of the chiral magnetic current in the static plasma in the small charge limit, with emphasis on the $\Delta$ dependence. Finally, we extend our discussion to an expanding plasma which is relevant for heavy-ion collision phenomenology. 

We solve the equations of motion numerically with a pseudospectral methods in the radial direction~\cite{Boyd1989ChebyshevAF}\footnote{See also appendix A of \cite{Grieninger:2020wsb} for a summary applied to a similar setup as our holographic model.} and a fourth order Runge Kutta scheme for the time evolution.

\subsection{Quasinormal modes}
\label{sec:qnm}
\begin{figure*}%[h!]
    \centering
    \subfloat{\includegraphics[scale=0.37]{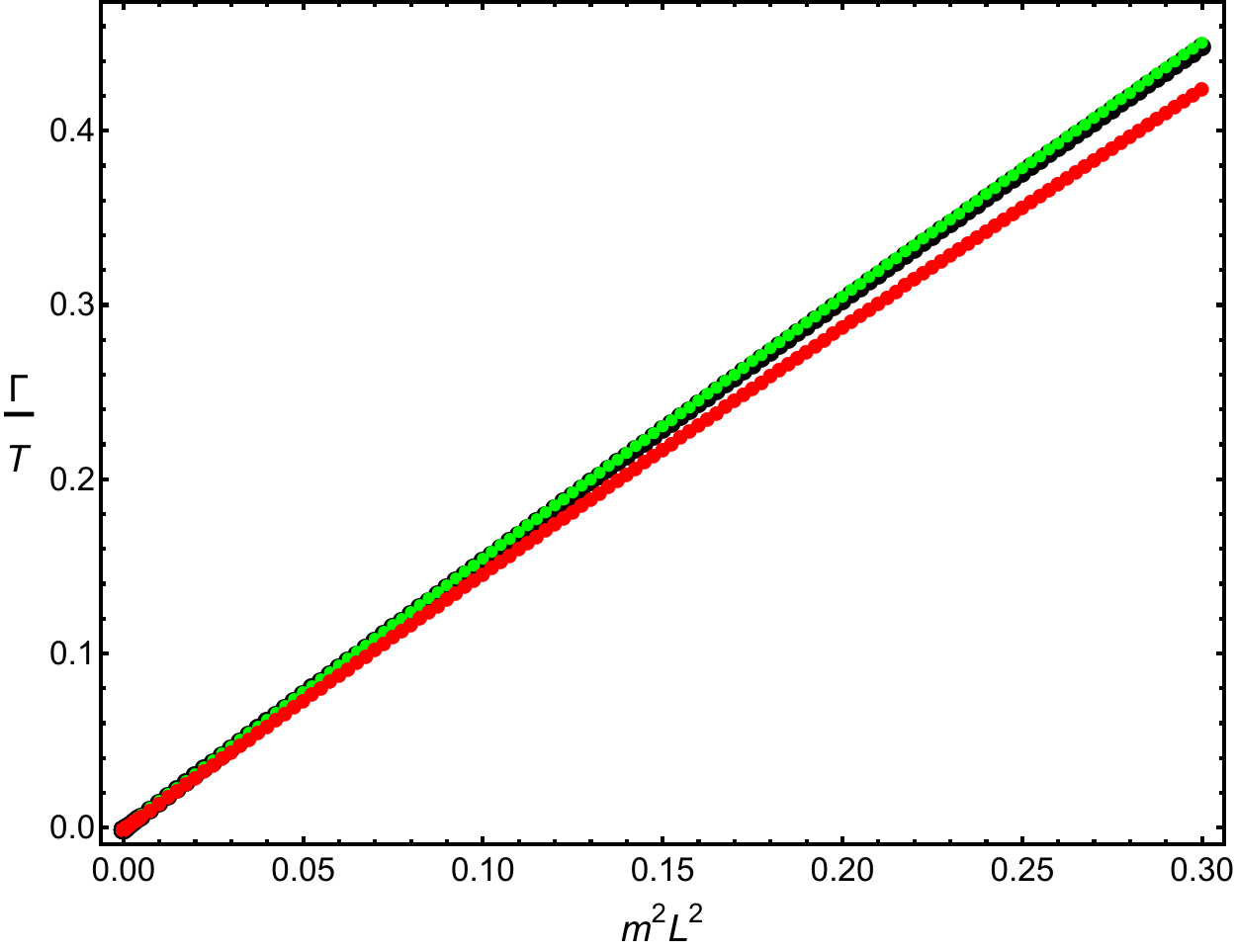}}
    \qquad
    \subfloat{\includegraphics[scale=0.39]{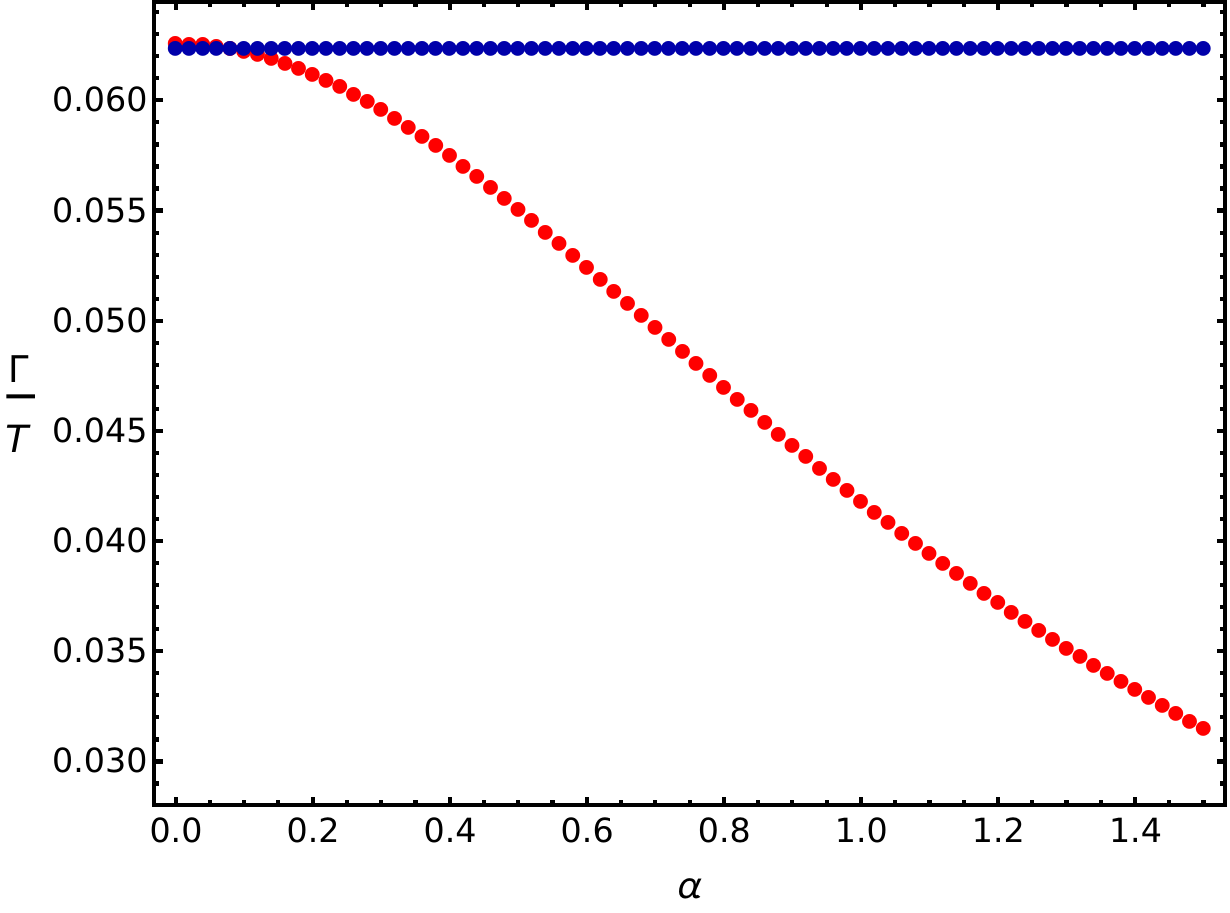}}
    \caption{\textbf{Left:} Dependence of the axial charge relaxation rate $\Gamma$ on the strength of the non-Abelian anomaly for two different values of the magnetic field. The black curve corresponds to $B/T^2=0.01$. The two curves for $\alpha=0$ and $\alpha=6/19$ are indistinguishable by eye. The green $(\alpha=0$) and red $(\alpha=6/19$) curve correspond to $B/T^2=2.96$. Note that the $\Gamma/T$ of the green curve is larger than that of the black curve. \textbf{Right:} Dependence of the axial charge relaxation rate on the strength of the Abelian anomaly $\alpha=6/19$ at fixed $m_s^2L^2=0.04$, and $B/T^2=0.02$ (blue) and $B/T^2=2.96$ (red). }
    \label{fig:qnm1}
\end{figure*}
The quasinormal modes (QNMs) for the unbroken $U(1)_V\times U(1)_A$ symmetry ($m_s=0$) were first computed in~\cite{Ammon:2016fru,Grieninger:2016xue}. In this section, we focus on the effect of finite $m_s$. At finite $m_s$, the $U(1)_A$ symmetry is explicitly broken and axial charge is no longer conserved. The chiral magnetic wave~\cite{Kharzeev:2010gd} is no longer a propagating sound wave but at small momenta (and in particular zero momentum) the dynamics is purely diffusive as shown in appendix A of \cite{Grieninger:2023wuq} or \cite{Jimenez-Alba:2014iia} (probe limit). The mode formerly associated with axial charge conservation acquires a gap in the imaginary part while the mode associated with vector charge conservation is still a (diffusive) hydrodynamic mode. Above a certain critical momentum the symmetry is restored and the chiral magnetic wave starts propagating. The gap $\omega_\text{gap}=-i\,\Gamma$ at zero momentum in the imaginary part determines the relaxation time of axial charge on which we will focus on in the following.

In the left panel of figure~\ref{fig:qnm1}, we show the dependence of the axial charge relaxation time on the strength of the non-Abelian anomaly governed by the mass $m_s$. As is evident, the nonconservation becomes stronger for increasing $m_s$ and axial charge has a shorter life time. Let us discuss the effect of $B$ in the relaxation time. The black curve corresponds to a small magnetic field ($B/T^2=0.01)$ and the results for $\alpha=0$ and $\alpha=6/19$ are not distinguishable by eye in this plot. The green and red curves correspond to a stronger magnetic field ($B/T^2=2.96$).  
The green curve corresponds to $\alpha=0$ and $\Gamma/T$ increases for stronger magnetic fields. For the red curve, $\alpha=6/19$ and $\Gamma/T$ decreases for the stronger magnetic field. In the right panel of figure~\ref{fig:qnm1}, we depict the dependence of the axial charge relaxation time on the Abelian anomaly at fixed $m_s^2L^2=0.04$, $B/T^2=0.02$ (blue) and $B/T^2=2.96$ (red). Contrasting to the left panel, increasing the strength of the Abelian anomaly ``protects'' axial charge for $B/T^2=2.96$ hence increasing its lifetime. The behavior of the blue curve is qualitatively similar but on a smaller overall scale. In \cite{Grieninger:2023wuq}, we show the dependence of the axial charge relaxation time on $B/T^2$ at fixed $\alpha$ and $m_sL$. For $\alpha\lesssim0.15$ increasing the magnetic field shortens the lifetime of axial charge. However, if $\alpha\gtrsim 0.15$ increasing $B/T^2$ protects axial charge and it is relaxing slower. In this work, we explicitly show that the real-time evolution follows the QNM prediction at late times (see appendix \ref{app:qnmslatetime}). Moreover, we will show that in the case of the expanding plasma (where the magnetic field decays with time) axial charge relaxation accelerates at late times (since weaker magnetic fields imply fast charge relaxation for $\alpha>0.15$).

\subsection{Static plasma dynamics}\label{sec:staticplasmadyn}
We first discuss the qualitative features of the chiral magnetic current and axial charge as we explore the parameter space $(B,T,\alpha,\Delta)$ and then provide new results concerning LHC and RHIC-like simulations. 

In order to solve the system of four equations \eqref{eq:eomstatic} we need to provide an initial state which satisfies the constraint equation. One unambiguous way of specifying the initial state is to choose two profiles $A(0,u)$ and $V(0,u)$ and then solve the constraint equation initially to find $\theta(0,u)$. The profiles ($A_t(0,u)$, $V_z(0,u)$) contain the information regarding the initial axial charge and chiral magnetic current which can be read off according to eqs. \eqref{eq:II.10} and \eqref{eq:II.10b}. Note that the equilibrium solution to eq. \eqref{eq:eomstatic} is trivial, which means that both axial charge and chiral magnetic current are identically zero. However, we know that axial charge may be generated in the initial stages of the collision and we may distinguish two qualitatively different scenarios: (A) Axial charge is generated before we start the holographic simulation or (B) Axial charge generations start simultaneously with the holographic simulation. In the first case, we should start with some nonzero amount of axial charge in the system and we implement that in the initial state by choosing $A_t(0,u) = n_5(0) u^{2+\Delta} $; in the second case we have vanishing initial charge and we assume that $\dot{q}_5(0) \neq 0$, so that $A_t(0,u) = \frac{2+\Delta}{3+\Delta} \Dot{q}_5(0)u^{3+\Delta} $, which triggers the generation of axial charge\footnote{We could have instead chosen a state where $\dot{q}_5(0)=0$ but the second (or $n$-th) derivative is different from zero and this would also lead to axial charge generation. The results for these alternative initial states are qualitatively similar to the ones presented here. The choice $\dot{q}_5(0)\neq0$ is further supported by the fact that axial charge obeys a first order differential equation. See also \cite{Fukushima:2010vw} for the equation that drives axial charge generation. }. In both cases we assume that there is no chiral magnetic current initially and we work with $V_z(0,u) = 0\,$.  It is useful to bear in mind that given a solution to eq. \eqref{eq:eomstatic}, we get a different solution rescaling ($A_t\,,V_z\,,\theta$) by the same amount. In particular this will rescale the value of $n_5(0)$ in (A) or the value of $\dot{q}_5(0)$ in (B). As a consequence, the qualitative features must be the same for simulations with different values of $n_5(0)$ or $\dot{q}_5(0)$ respectively. Nonetheless, we stress that the result is only valid for small values of $n_5\,$.

\subsection*{Parameter space: Qualitative features.}

The chiral magnetic current in a static plasma without the St\"uckelberg field (i.e. $m_s=\Delta=0$) was studied in \cite{Ghosh:2021naw}. There, we discussed that a physically sensitive value for the anomaly coefficient is $\alpha=\frac{6}{19}$ (see also the discussion leading to eq. \eqref{eq:newton}). However, we observed that qualitative differences are amplified for bigger values of $\alpha$ and so we fix $\alpha=1.5$ for this section.
Finally, we display simulations where we vary the dimensionless parameter $B/T^2$, ranging from $\sim 1$ to $\sim 10$. Since our solutions come from a linear system, we fix rescalings by normalizing both vector current and axial charge to the peak value of axial charge.  

The results for $\alpha=1.5$ are shown in figure \ref{fig:adim15} in the appendix \ref{app:parameterscanstatic}. We find that higher values of $\Delta$ result into faster dissipation, since $\Delta$ measures the nonconservation of the axial charge. Moreover, dissipation is more significant for lower magnetic fields as we already expected from the QNM results of the previous section. The chiral magnetic current and axial charge display oscillatory behavior, which is more prominent as the magnetic field is increased. For the chiral magnetic current the presence/absence of oscillations is independent of the value of $\Delta$ while for the axial charge they are amplified as we increase $\Delta$.

Let us now briefly discuss the results for $\alpha=\frac{6}{19}$.
As expected, higher values of $\Delta$ yield faster dissipation. The oscillatory behavior is absent even for the strongest magnetic field. Actually, in \cite{Ammon:2016fru,Grieninger:2016xue} it was shown that the quasinormal modes are controlled by $\alpha B$, so the small value of $\alpha$ puts us into the parameter space where oscillations are absent (for the considered magnetic fields). As for the initial response, it can be checked numerically that the chiral magnetic current reacts (slightly) faster (in dimensionless units) when the magnetic field increases.

In table \ref{tab:static1}, we characterize the time it takes for the chiral magnetic current and axial charge density to reach their maximum value. At fixed $\Delta$, then we see that the chiral magnetic current peaks faster if we increase the strength of the magnetic field while the opposite is true for the axial charge density which peaks slower at larger values of the magnetic field. If we fix $B/T^2$ and increase $\Delta$ we see that the chiral magnetic current and the axial charge density peak faster.

Finally, in appendix \ref{app:qnmslatetime}, we explicitly demonstrate that the exponential late time falloffs match the QNM frequencies as expected.

\begin{table*}%[h!]
\begin{tabular}{cccccccccccccccc}
$\Delta$ & \multicolumn{5}{|c}{$0.001$} & \multicolumn{5}{|c}{$0.05$} & \multicolumn{5}{|c}{$0.29$} \\
$\frac{B}{T^2}\,\,$ & \multicolumn{1}{|c}{$0.99\,\,$} &  $2.98\,\,$  &  $5.04\,\,$  & $7.19\,\,$  &  $10.72\,\,$ &   \multicolumn{1}{|c}{$0.99\,\,$} &  $2.98\,\,$  &  $5.04\,\,$  & $7.19\,\,$  &  $10.72\,\,$   &    \multicolumn{1}{|c}{$0.99\,\,$} &  $2.98\,\,$  &  $5.04\,\,$  & $7.19\,\,$  &  $10.72\,\,$   \\
\hline
\hline
$\overline{v}_{J_\text{max}^A} $ & \multicolumn{1}{|c}{$2.3$}   & $2.2$    &  $2.1$  & $2.0$  & $1.8$  &  \multicolumn{1}{|c}{$2.2$}  & $2.1$    & $2.0$   & $1.9$  &  $1.7$ &  \multicolumn{1}{|c}{$1.9$}  &  $1.9$  &  $1.8$  & $1.7$  & $1.6$  \\
$\overline{v}_{J_\text{max}^B}$ &  \multicolumn{1}{|c}{$2.5$}  &  $2.4$   &  $2.3$  & $2.1$  & $2.0$  &  \multicolumn{1}{|c}{$2.3$}  &  $2.3$  &  $2.2$  & $2.1$  & $2.0$  &  \multicolumn{1}{|c}{$2.1$}  &  $2.1$  & $2.0$   & $2.0$  & $1.9$  \\
 $\overline{v}_{J_{5,\text{max}}^B}$&  \multicolumn{1}{|c}{$1.75$}  & $1.76$    & $1.77$    & $1.78$  & $1.80$  & \multicolumn{1}{|c}{$1.63$} &  $1.63$  &    $1.65$ & $1.66$  & $1.70$  & \multicolumn{1}{|c}{$1.38$}    & $1.39$    & $1.41$    & $1.44$  &  $1.50$\\
 \hline
 \hline
\end{tabular}
\caption{Time (in dimensionless units $\overline{v}\equiv v\epsilon_L^{1/4}$) at which chiral magnetic current $J$ finds its peak value for initial states $A$ and $B$ and axial charge $J_5$ finds its peak value for initial state $B$. We fixed $\alpha = \frac{6}{19}\,$.}\label{tab:static1}
\end{table*}
%\end{widetext}
\subsection*{Results for RHIC and LHC parameters.}

We now perform simulations with parameters relevant for the quark-gluon plasma. We shall work with the estimated value of $\alpha=\frac{6}{19}$ \cite{Ghosh:2021naw}. The value of $\Delta$ is not known but it is expected to be small. We take here a conservative approach and display results for two distinct values: $\Delta(m_s^2=1/499) \simeq0.001\,$ and $\Delta(\sqrt{2}/3)\simeq 0.11 \,$. 
As initial states, we again use: (A) having a finite axial charge initially and zero chiral magnetic current current: $V_z(0,u)=0\,$, $A_t(0,u)= n_5(0) u^{2+\Delta}$ and (B) having initially zero axial charge and chiral magnetic current but nonzero time derivative for the axial charge: $V_z(0,u)=0\,$, $A_t(0,u)= \frac{2+\Delta}{3+\Delta} \Dot{q}_5(0)u^{3+\Delta}\,$.

In order to choose sensitive parameters for the plasma conditions at both RHIC and LHC we follow \cite{Shi:2017cpu}, which gives a centrality dependence of magnetic field peak and axial charge density normalized to entropy density for $Au-Au$ collisions at $\sqrt{s}=200 \,\si{GeV}$. We reproduce it in table \ref{tab:1} for RHIC parameters and in table \ref{tab:2} for LHC parameters\footnote{We thank Shuzhe Shi for providing us with the magnetic field for LHC collisions, in particular $Pb-Pb$ collisions at $\sqrt{s}=5.02\,\si{TeV}$, (estimations done with the Optical Glauber Model) at different centralities as well as the temperatures for both LHC and RHIC. }. We proceed now to elucidate the meaning of $B_\text{max}$ and $T_0$. 

In \cite{Shi:2017cpu} the time evolution of the magnetic field is parametrized as
\begin{align}\label{eq:timeB}
B(\tau)&=\frac{B_\text{max}}{1+\tau^2/\tau_B^2},
\end{align}
%\noindent
where $\tau_B$ is the lifetime of the magnetic field. As discussed in \cite{Ghosh:2021naw}, we may take $\tau_B^\text{RHIC}=0.6 \,\si{fm/c}$ and $\tau_B^\text{LHC} = 0.02 \,\si{fm/c}\,$. The value of temperature corresponds to the equilibration time, which may be taken to $\tau_0=0.6 \,\si{fm/c}$ for both cases. At late times, in the Bjorken regime, temperature evolves according to 
\begin{equation}\label{eq:timeT}
T(\tau) = T_0 \left(\frac{\tau_0}{\tau}\right)^{1/3}.
\end{equation}
\noindent
Values of axial charge $n_5$ at different initial temperature may be obtained using the scaling relation 
\begin{equation}
(n_5/s)_\text{LHC}=\left(\frac{T_\text{RHIC}}{T_\text{LHC}}\right)^{3/2}(n_5/s)_\text{RHIC}\,.
\end{equation}
Notice that the value $B_\text{max}$ is obtained at $\tau=0$, whereas $T_0$ is obtained at $\tau_0$. In order to obtain a consistent picture we use \ref{eq:timeB} and \ref{eq:timeT} to obtain both $B$ and $T$ at some intermediate time between the plasma formation\footnote{See \cite{Feng:2016srp} for an estimation of the plasma formation time.} and the equilibration time. In particular we choose $\tau_{sim}^\text{RHIC} = 0.3 \,\si{fm/c}$ and $\tau_{sim}^\text{LHC}=0.1\,\si{fm/c}$. This finally reproduces the values $B_{sim}$ and $T_{sim}$ found in both tables \ref{tab:1} and \ref{tab:2}. Lastly, the value of axial charge is taken to be the peak value. We obtain multiplying $n_5/s$ by the black hole entropy, which has been already matched to the expected entropy of the plasma \cite{Ghosh:2021naw}. 

We stress that in this section the plasma is nonexpanding and neither temperature nor magnetic field evolve in this setup. Similarly, our ansatz is homogeneous and it is therefore not possible to simulate genuine off-centered collisions. The values displayed in both tables are meant to give representative parameters for the simulation and may serve as a guide for future studies with more refined holographic setups. In the next section, we present phenomenologically more realistic simulations with an expanding plasma where temperature and magnetic field decrease over time. 

\begin{table}[h!]
\begin{tabular}{lcccc}
\hline
Centrality bin & $10-20\%\,\,\,\,$ & $20-30\%\,\,\,\,$ & $30-40\%\,\,\,\,$ & $40-50\%\,\,\,\,$ \\
\hline
\hline
$(n_5/s)_0$ & $0.065$ & $0.078$ & $0.095$ & $0.119$ \\
\hline
$T_0\,(\si{GeV})$ & $0.341$ & $0.329$ & $0.312$ & $0.294$\\
\hline
$eB_\text{max}(m_\pi^2)$ & $2.34$ & $3.1$ & $3.62$ & $4.01$ \\
\hline
\hline
$T_{sim}\,(\si{GeV})$ & $0.429$ & $0.414$ & $0.393$ & $0.370$ \\
\hline
$eB_{sim}(m_\pi^2)$ & $1.87$ & $2.48$ & $2.90$ & $3.20$ \\
\hline
\end{tabular}
\caption{Data for $Au-Au$ collisions in RHIC at energy $\sqrt{s}=200\,\si{GeV}$. The first three rows are taken from~\cite{Shi:2017cpu}. The subscript \textit{sim} indicates the value at the initial time of our simulation $\tau_{sim}^\text{RHIC} = 0.3 \,\si{fm/c}$.}
\label{tab:1}
\end{table}
%\begin{widetext}
\begin{table}[h!]
\begin{tabular}{lcccc}
\hline
Centrality bin & $10-20\%\,\,\,\,$ & $20-30\%\,\,\,\,$ & $30-40\%\,\,\,\,$ & $40-50\%\,\,\,\,$ \\
\hline
\hline
$(n_5/s)_0$ & $0.039$ & $0.045$ & $0.059$ & $0.075$ \\
\hline
$T_0\,(\si{GeV})$ & $0.48$ & $0.47$ & $0.43$ & $0.40$\\
\hline
$eB_\text{max}(m_\pi^2)$ & $59.2$ & $78.5$ & $91.7$ & $101.6$ \\
\hline
\hline
$T_{sim}\,(\si{GeV})$ & $0.87$ & $0.85$ & $0.78$ & $0.73$\\
\hline
$eB_{sim}(m_\pi^2)$ & $2.28$ & $3.02$ & $3.53$ & $3.91$ \\
\hline
\end{tabular}
\caption{Data for $Pb-Pb$ collisions in LHC at energy $\sqrt{s}=5.02\,\si{TeV}$. The subscript \textit{sim} indicates the value at the initial time of our simulation $\tau_{sim}^{LHC} = 0.1 \,\si{fm/c}$.}
\label{tab:2}
\end{table}%\end{widetext}
\begin{figure*}%[h!]
	\centering
	\includegraphics[width=0.48\linewidth]{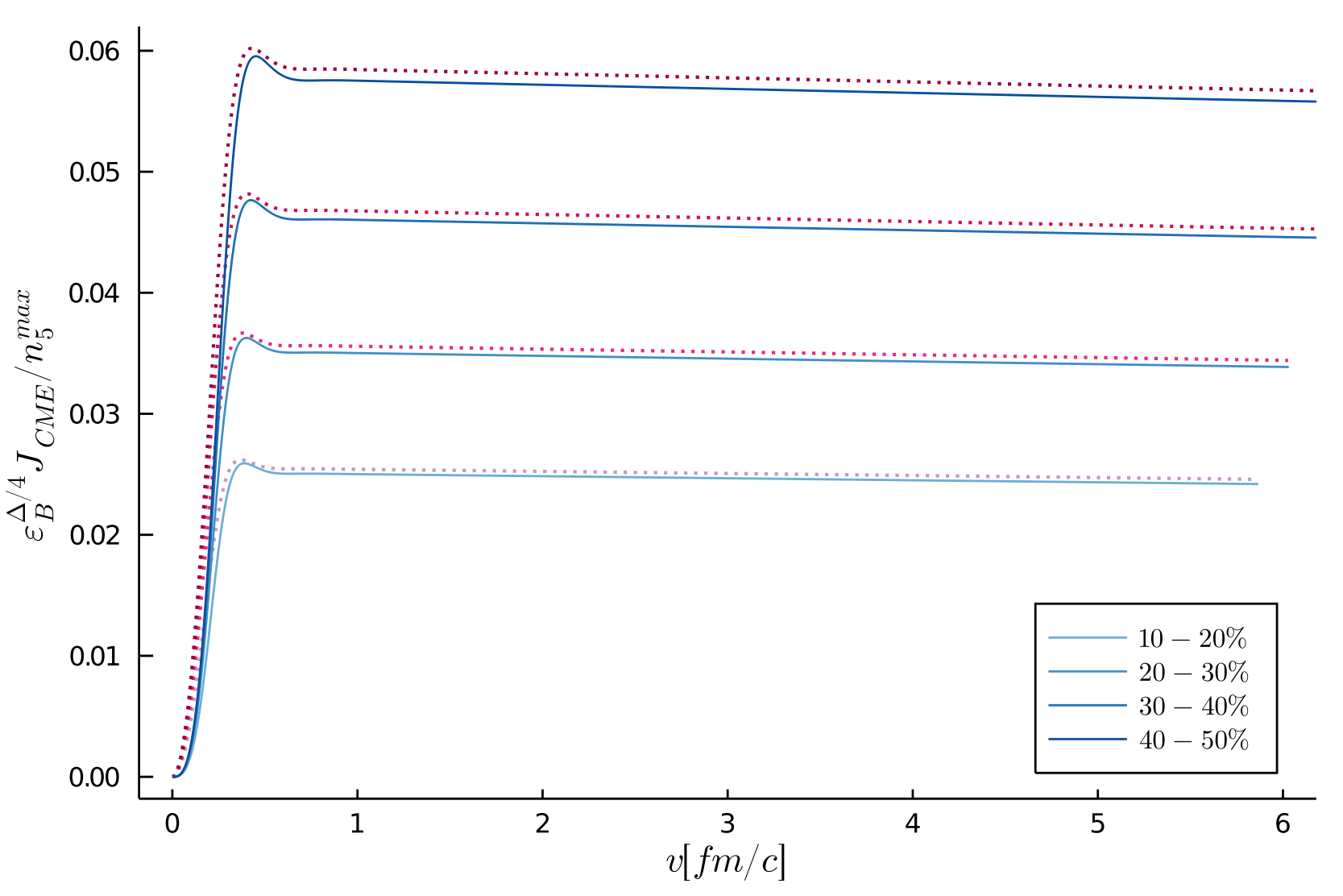}
	%\qquad\vspace{0.4cm}
	%\subfloat{
 \includegraphics[width=0.48\linewidth]{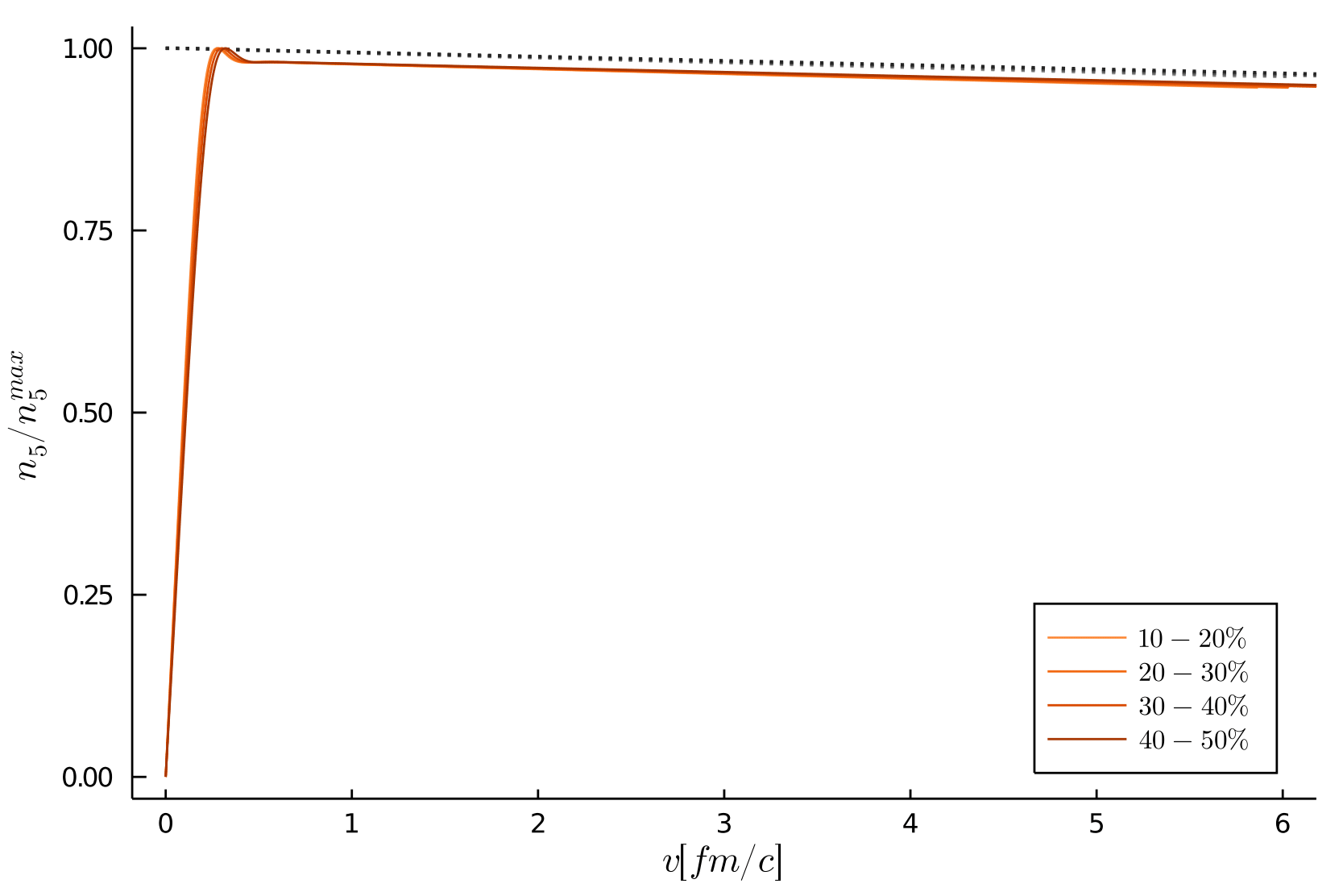}%}
	\qquad 
	%\subfloat{
 \includegraphics[width=0.48\linewidth]{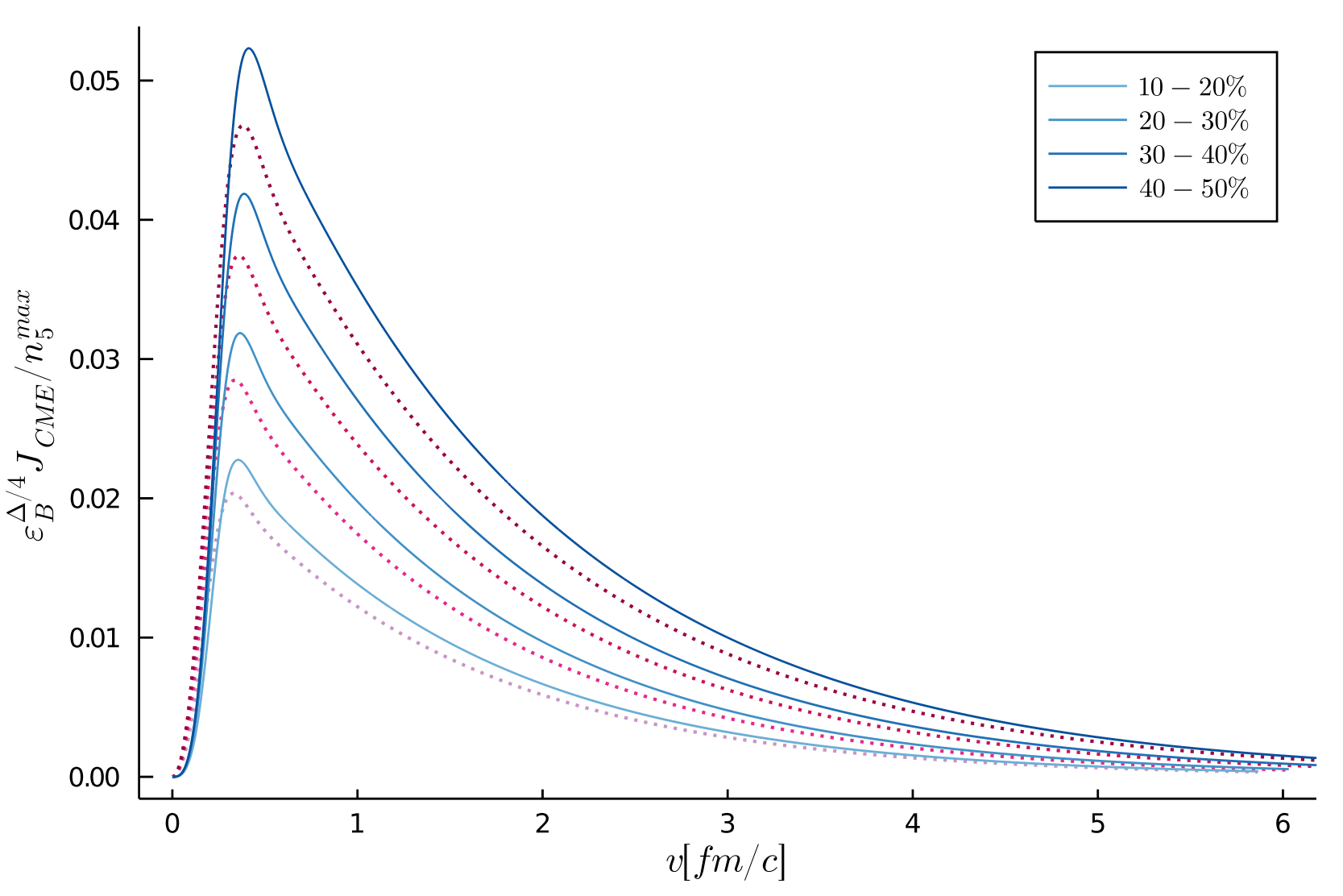}%}
	%\qquad\vspace{0.4cm}
	%\subfloat{
 \includegraphics[width=0.48\linewidth]{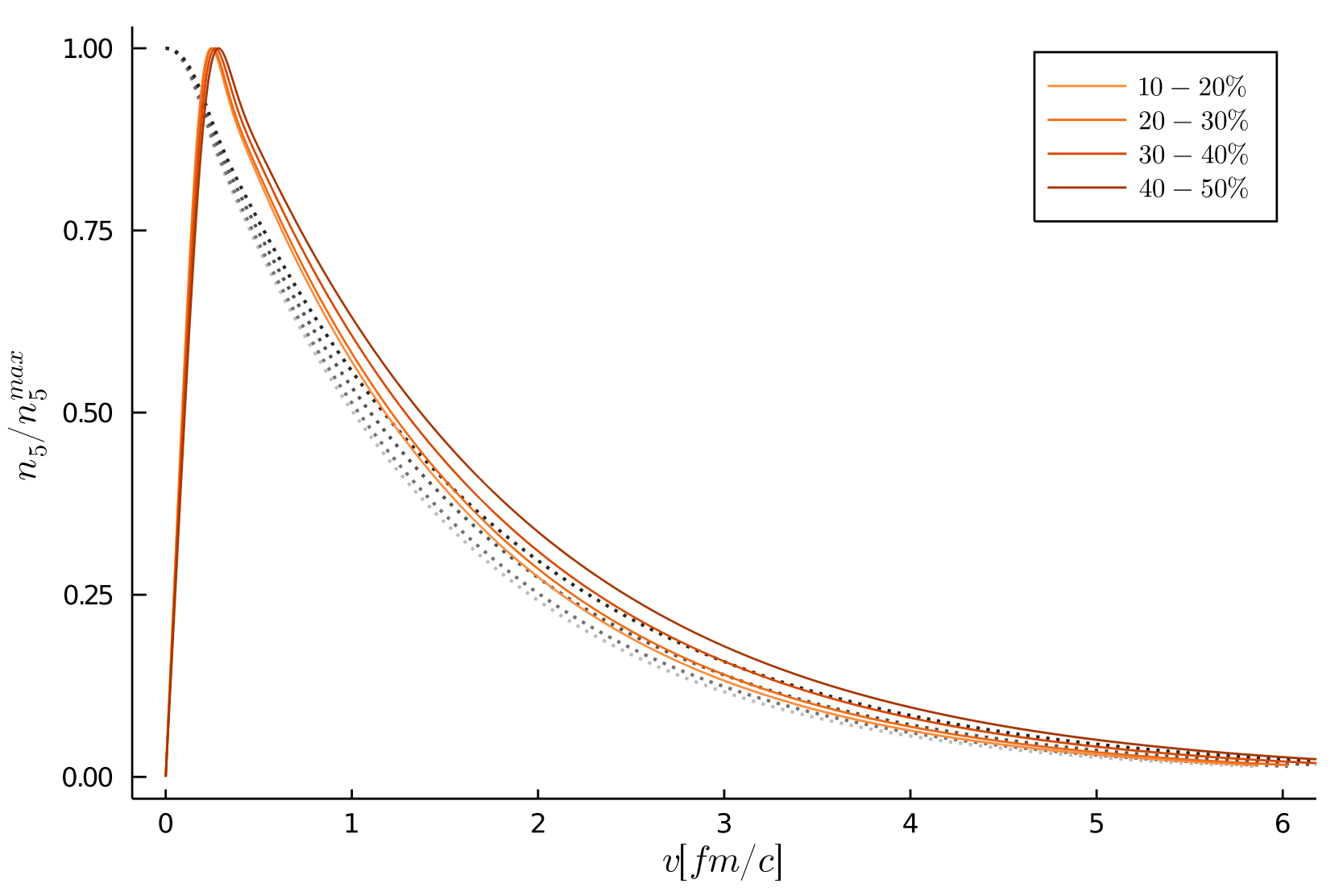}%}
	\caption{Solid lines correspond to initial state (A) whereas dashed lines correspond to state (B). (\textbf{Left:}) Chiral magnetic effect and (\textbf{Right:}) axial charge for simulations with RHIC-like parameters as a function of centrality. We set $\alpha=\frac{6}{19}$ as well as $\Delta=0.001$ (top) and $\Delta=0.11$, respectively.  The labeling refers to data in table \ref{tab:1} applies to both solid and dashed lines in a correlated manner. }
	\label{fig:RHICn}
\end{figure*}

The results for RHIC and LHC-like simulations are displayed in figures \ref{fig:RHICn} and \ref{fig:LHCn} (appendix \ref{app:parameterscanstatic}), respectively. In both cases increasing $\Delta$ leads to stronger charge dissipation. The vector current obtained from states (A) and (B) are roughly indistinguishable for small values of $\Delta$, whereas for higher values the overall amplitude differs up to a factor of $3/4\,$. After the peak, axial charge decays according to the the quasinormal mode behavior outlined in the previous section. In both figures the parameters corresponding to bigger centralities result into larger chiral magnetic current (compared to the peak value of axial charge).\footnote{We stress that the peak value of axial charge also depends on centrality. According to the data in tables \ref{tab:1} and \ref{tab:2}, the peak value of axial charge also increases as a function of centrality.} The reason is that the ratio $B/T^2$ is also higher for these cases. The vector current and axial charge in LHC-like simulations build up faster than in RHIC-like simulations. From the previous section we know that the buildup time is roughly constant in dimensionless units. Turning the result into dimensionful units accounts for the difference at different energies.  

Remarkably, the CME obtained for LHC-like parameters is smaller by a factor of 3 compared to RHIC-like simulations in dimensionless units. In other words, we would need $n_5^\text{LHC}\simeq 3 n_5^\text{RHIC}$ for both signals to be roughly equal in amplitude.
The explanation lies in the much shorter lifetime of the magnetic field at LHC as compared to RHIC. One could argue that choosing an earlier time for the LHC simulation would result into a higher magnetic field and consequently larger chiral magnetic current, however, as discussed previously, it is not clear that any plasma has formed before $\tau=0.1 \,\si{fm/c}$ at LHC.  Indeed, estimates of the axial charge in the literature \cite{Cartwright:2021maz,PhysRevC.98.014911} seem to give $n_5^\text{LHC}\simeq 3 n_5^\text{RHIC}$.

The precise value of the chiral magnetic current is to be taken as crude estimation, since (among other things) the magnetic field is static in this setup. However, this is again in line with the statement that the chiral magnetic effect is favored at RHIC, a similar conclusion to \cite{Ghosh:2021naw}. This result is seemingly different from what we will obtain in the next section (and what was also obtained in \cite{Cartwright:2021maz}) where lower temperatures lead to less CME signal. The difference in the results are the different treatment in the lifetime of the magnetic field and its decay with time (in one simulation we fix the initial data assuming a decay of $\sim 1/(1+\tau^2/\tau_B^2)$ and then keep $B$ static. In the expanding plasma simulation the magnetic field decays as $\sim 1/\tau$ and thus the two simulations are not really comparable).

\subsection{Expanding plasma dynamics: Initial parameters}\label{sec:expandingone}
\noindent
In order to perform simulations for phenomenologically relevant parameters, we connect our holographic simulation to hydrodynamics. We stress that the following equations and approximations are only used to approximate initial conditions for our holographic simulation. The holographic computation is valid beyond Bjorken hydrodynamics and the time evolution of the (holographic) energy density is determined by the bulk equations of motion. We aim to derive a flow parameter at late times which we can use to express our quantities in dimensionless units. We normalize our quantities to the late time flow parameter and also \textit{a posteriori} adjust the initial conditions that lead to the desired late time behavior. 
Neglecting the magnetization of the plasma, the evolution of the energy density in the Bjorken regime reduces to~\cite{Critelli:2018osu,Roy:2015kma,Pu:2016ayh,Rougemont:2023gfz,Cartwright:2022hlg}
\begin{equation}
    \partial_\tau\left(\epsilon+\frac{b(\tau)^2}{2}\right)+\frac 43\frac{\epsilon}{\tau}-\frac{4}{3}\frac{\eta}{\tau^2}+\frac{b(\tau)^2}{\tau}=0.\label{eq:hydroeps}
\end{equation}
In a (strong) magnetic field, the viscosity $\eta$ is of course not a scalar quantity but a tensor~\cite{Hernandez:2017mch,Ammon:2020rvg}. However, for simplicity we treat it as a scalar which is valid for $B\ll T^2$.
Using $\eta/s=1/(4\pi)\,$, $4/3\epsilon=4p=sT\approx 4/3cT^4\,$,\footnote{These equations hold for conformal theories at zero charge -- recall that we neglect the backreaction of the charge density onto the evolution of the energy momentum tensor -- and if we neglect the susceptibilities that appear in the equation of state due to the magnetic field, which are negligible at late times. Moreover, the constant $c$ defines the normalization of energy density to temperature.} and $b(\tau)=B/\tau$, we find
\begin{equation}
    \partial_\tau\epsilon+\frac43\frac \epsilon\tau=\frac{4}{9\, c\,\pi\, \tau^2}\epsilon^{3/4}
\end{equation}
with solution
\begin{equation}\label{eq:dynamicseps}
    \epsilon(\tau)=\frac{\left(c^{1/4}-6 \pi\, c_1\, \tau^{2/3}\right)^4}{1296 \pi^4\,\tau^4},%\epsilon_0\left(\frac{\tau_0} {\tau}\right)^{4/3}e^{-\frac{4 (\tau-\tau_0)}{9 \pi  T_0\,\tau\tau_0}}.
\end{equation}
which scales like $\epsilon\sim c_1^4/\tau^{4/3}+\ldots$ at late times.
The constant $c_1$ is determined by connecting to the expression known from ideal Bjorken hydrodynamics $\epsilon\sim \epsilon_\infty\left(\frac{\tau_\infty} {\tau}\right)^{4/3}+\mathcal O(1/\tau^2)$, i.e. $c_1=\epsilon_\infty^{1/4} \tau_\infty^{1/3}$.

We choose our initial parameters as follows. From lattice QCD~\cite{Karsch:2000ps} or from the equation of state for conformal fluids, we know that the energy density is related to the temperature by $\epsilon=3p=3/4\,sT=\frac{19\pi^2}{16}T^4$ for temperatures around $T=300-\SI{350}{\MeV}$. This fixes the constant $c$ in eq. \eqref{eq:dynamicseps} as $c=19\pi^2/16$. In RHIC collisions with beam energy $\sqrt{s}=\SI{200}{\GeV}$ the parameters commonly used in hydrodynamic simulations are $B=m_\pi^2$ for the magnetic field strength and $T_0=\SI{300}{\MeV}$ for temperature. Together with initial time $\tau_0\sim 0.6\,\text{fm}$~\cite{Shi:2018izg} the initial conditions yield the dimensionless quantities $\epsilon_B/B^2\approx 247$ and $\tau_0\epsilon_0^{1/4}\approx 1.69$. In addition to this, we consider a slightly lower and higher beam energy as indicated in table \ref{tab:iniexp}.
In our holographic simulations, we can go beyond (Bjorken) hydrodynamics and start our simulations at an earlier initial time.

\begin{table}
\begin{center}
\begin{tabular}{l| c c c }
\hline &&&\\[-8.5pt]
Beam energy: &$\sqrt{s}=\SI{175}{\GeV}$&$\sqrt{s}=\SI{200}{\GeV}$&$\sqrt{s}=\SI{250}{\GeV}$ 		 \\ \addlinespace\addlinespace
\hline\hline
$T_0$&\SI{289}{\MeV}&$\SI{300}{\MeV}$&$\SI{316}{\MeV}$  \\
$B$ &$0.875m_\pi^2$&$m_\pi^2$&$1.25m_\pi^2$  \\
$n^\text{max}_5/s^{1+\Delta/3}$&$0.875\times 0.065$&$0.065$&$1.25\times 0.065$\\
 $\epsilon_B/B^2$& 279&247&194\\
 $\tau_0\epsilon_0^{1/4}$&1.63&1.69&1.78\\
 \addlinespace
\hline 
\end{tabular}
\caption{Initial data expanding plasma simulation where we assumed on the hydro side that the values for $B$ and $T$ are for $\tau_\text{ini}=0.6\,\text{fm}$.}
\label{tab:iniexp}
\end{center}
\end{table}
\begin{figure*}%[h!]
    \centering
    \includegraphics[width=0.49\linewidth]{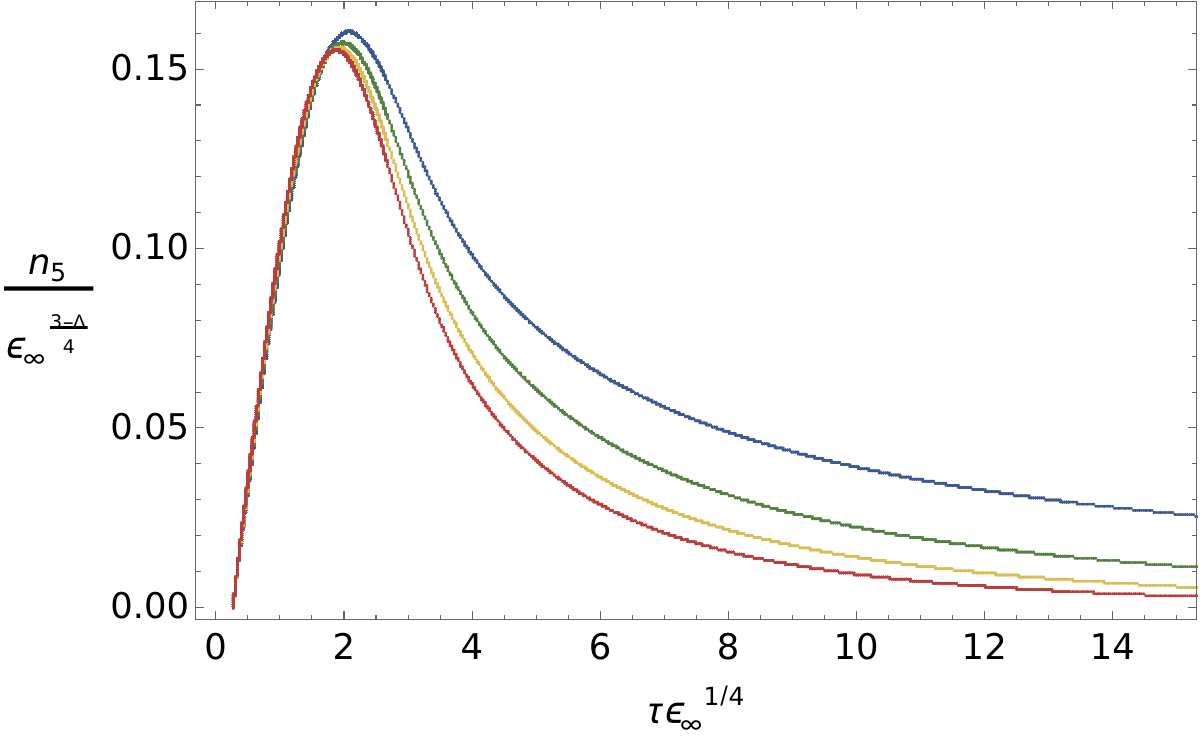}\ 
\includegraphics[width=0.49\linewidth]{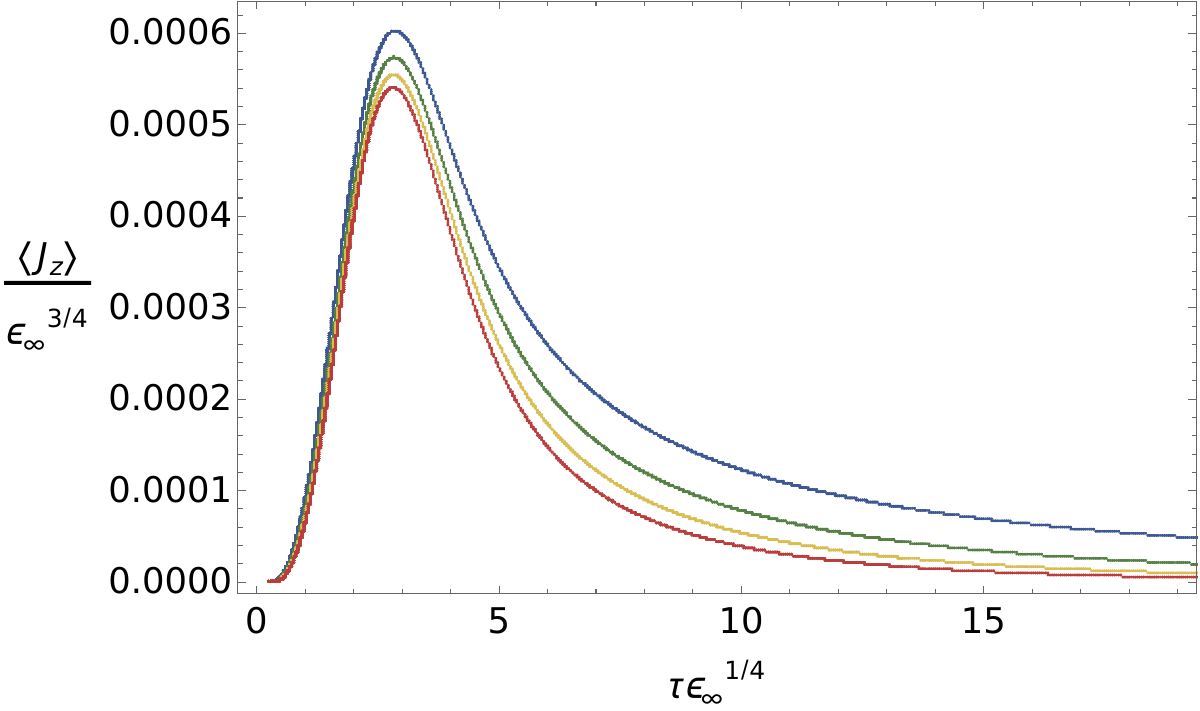}
    \caption{Axial charge density (left) and chiral magnetic current (right) as a function of time corresponding to the $\sqrt{s}=200\,\si{\GeV}$ initial conditions. The coupling $m_s$ increases from blue ($\Delta=1.25\times 10^{-7}$) to red $(\Delta=0.3$). }
    \label{fig:exp1}
\end{figure*}
\begin{figure*}%[h!]
    \centering
    \subfloat{\includegraphics[scale=0.4]{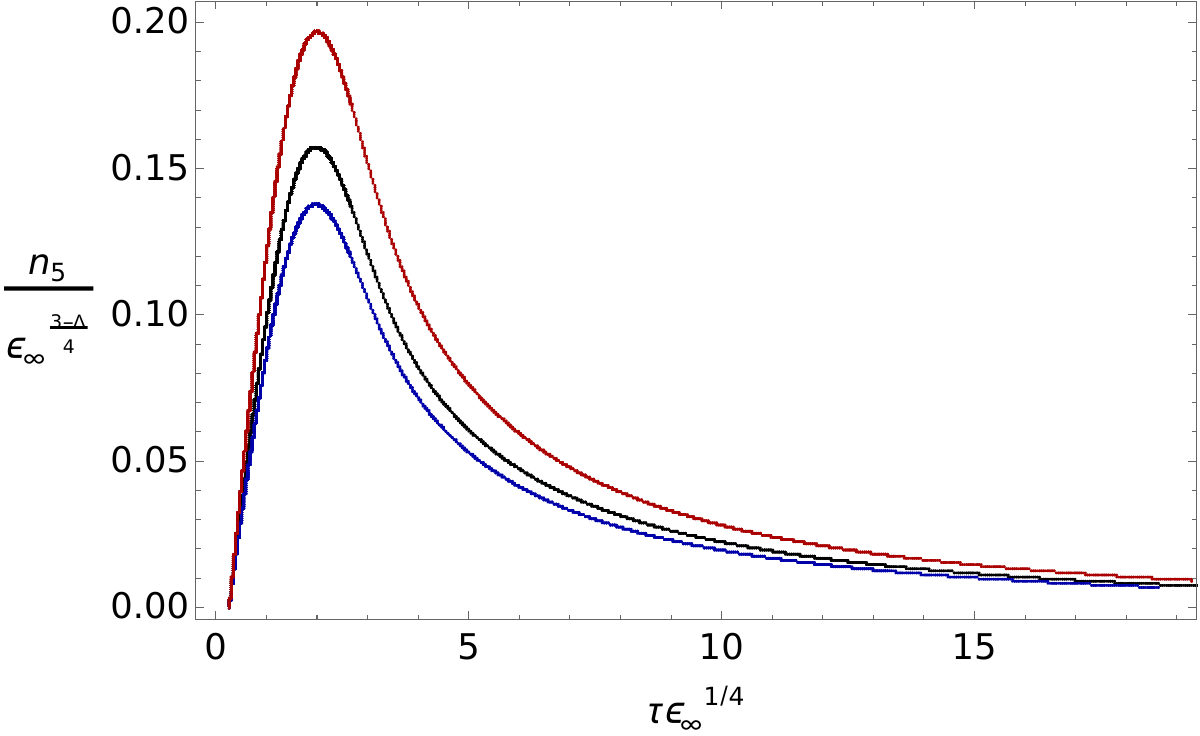}}
    \
    \subfloat{\includegraphics[scale=0.4]{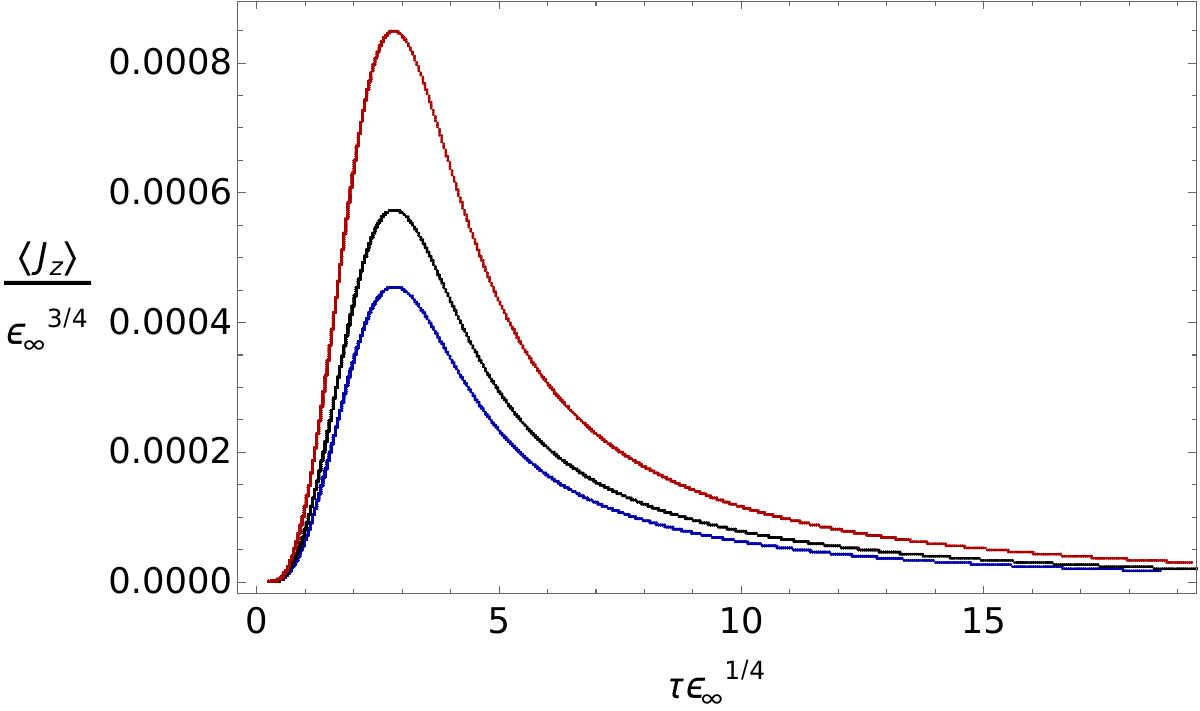}}
    \caption{Axial charge density (left) and chiral magnetic current (right) as a function of time corresponding to the $\sqrt{s}=250\,\si{\GeV}$ (red), $\sqrt{s}=200\,\si{\GeV}$ (black) and $\sqrt{s}=150\,\si{\GeV}$ (blue) initial conditions. The coupling $\Delta=0.118$ is fixed.}
    \label{fig:exp2}
\end{figure*}

In order to connect to the hydrodynamic simulations, we adjust our initial parameter according to the following procedure:
Starting from a fixed initial state with vanishing axial charge, chiral magnetic current and dynamical pressure at a fixed initial time $(\tau L=0.2)$ we determine $\varepsilon_\infty$ by fitting the late time behavior to eq.~\eqref{eq:dynamicseps}. We then adjust the initial magnetic field $B$ and energy density $\varepsilon$ on the holographic side until we find an $\varepsilon_\infty$ that satisfies the two dimensionless ratios $\varepsilon_\infty/B^2$ and $\tau \varepsilon^{1/4}_\infty$ as indicated in table \ref{tab:iniexp}. Fixing the initial data in this way is similar to the procedure outlined in~\cite{Chesler:2009cy}. By determining the initial data by adjusting $\varepsilon_\infty$ in eq.~\eqref{eq:dynamicseps}, we connect to a Bjorken hydro simulation that runs through our desired initial values (even though the energy density in our simulation does not go through those parameter pairs since the holographic computation is beyond Bjorken hydrodynamics at early times). It has also the advantage that we can normalize our quantities to time-independent quantities.

\subsection{Expanding plasma: Numerical results}\label{sec:expandingtwo}
In figure \ref{fig:exp1}, we depict the simulation for the axial charge and chiral magnetic current corresponding to the the $\sqrt{s}=200\,\si{\GeV}$ initial data outlined in table~\ref{tab:iniexp}. The mass $m_s$ that governs the nonconservation of axial charge increases from black to red resulting in a faster decaying axial charge and smaller CME signal. At late times and $\Delta=0$, we expect the axial charge density and the magnetic field to decrease like $\sim 1/\tau$ due to dilution. At late times and small $B/T^2$, the axial susceptibility we can use $n_5\sim \chi_5\,\mu_5\sim z_h^{-2} \mu_5+\mathcal O(B/T^2)$ (and $T\sim z_h^{-1}\sim\tau^{-1/3}$ at late times). Hence, the chiral magnetic current should fall off as $\langle J_\text{CME}\rangle\sim \alpha B n_5/\chi_5\sim 1/\tau^{4/3}$. At finite $m_s$ the falloffs of $n_5$ and $\langle J_\text{CME}\rangle$ are accelerated due to the nonconservation. 

\begin{figure*}%[h!]
    \centering
  \subfloat{\includegraphics[scale=0.4]{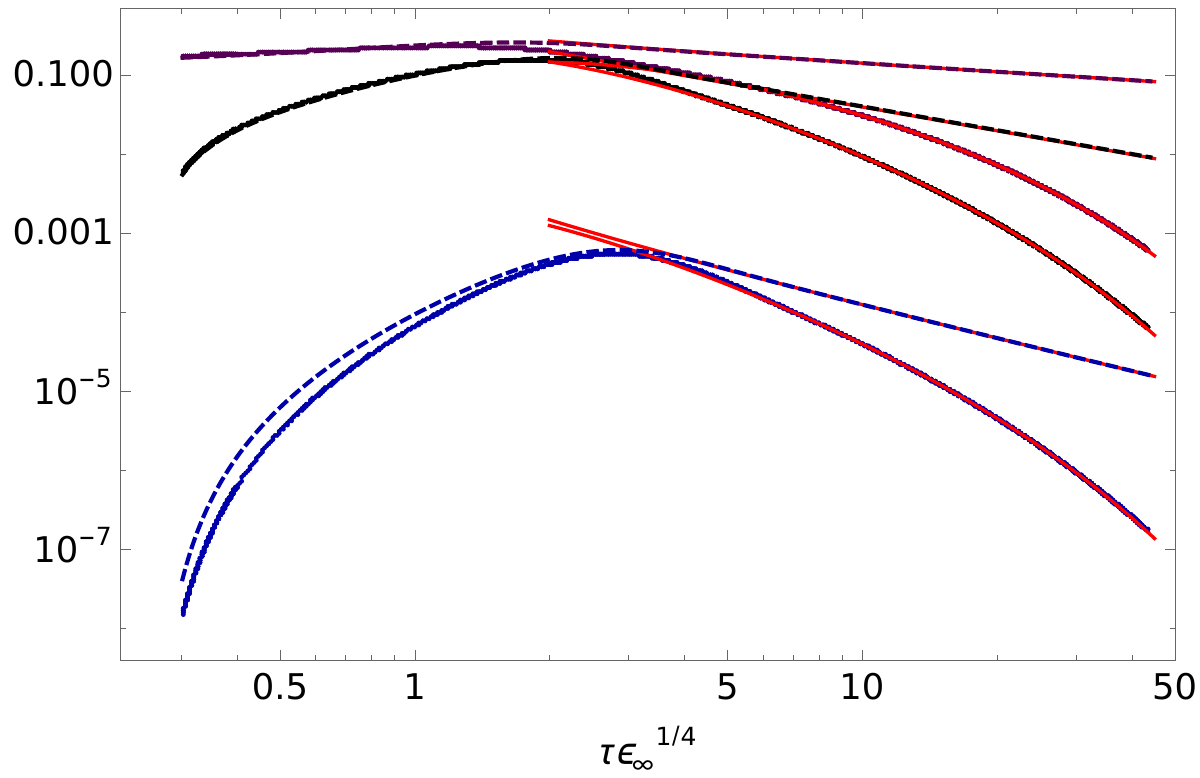}}
   \  \subfloat{\includegraphics[scale=0.4]{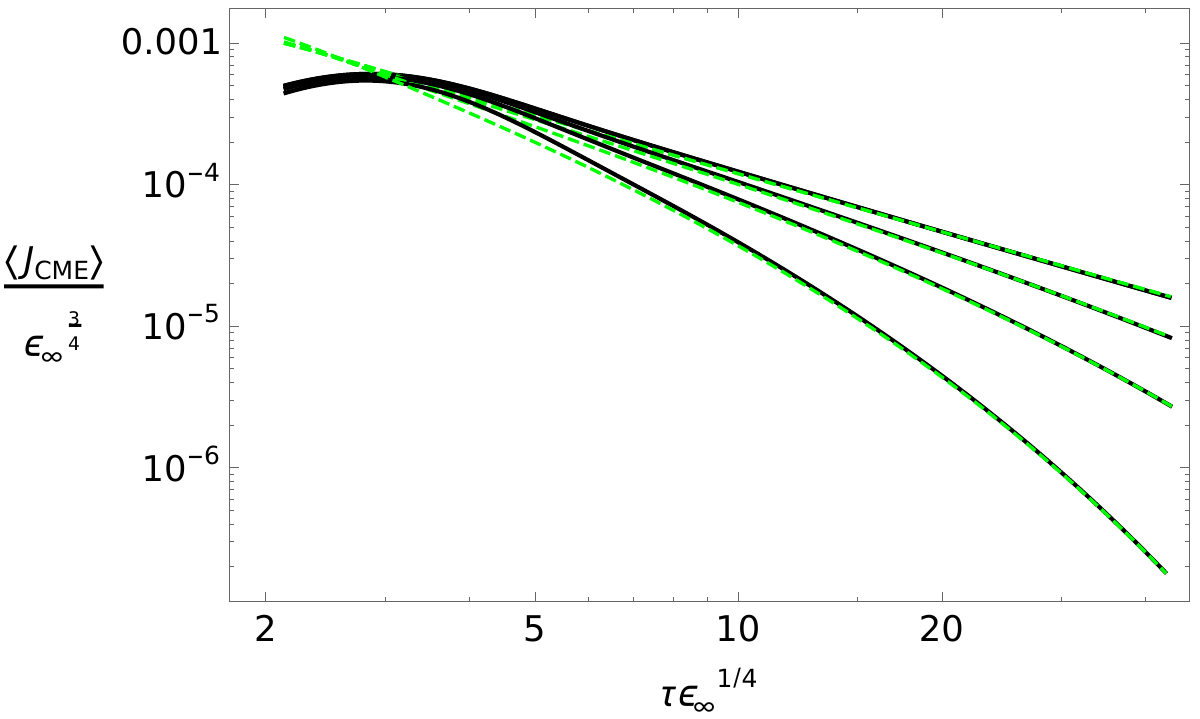}}
   \caption{Left: Time evolution of $\langle J_\text{CME}\rangle/\epsilon_\infty^{3/4}$ (blue), $n_5/e/\epsilon_\infty^{3/4+\Delta/4}$ (black) and $A_v(1)$ (purple). The dashed lines correspond to $\Delta=1.25 10^{-7}$ and the solid lines to $\Delta=0.3$. The red lines are the fits outlined in the main text. Right: Double logarithmic plot depicting the late time behavior of the chiral magnetic current for different $\Delta$ (black lines). The green dashed lines are formula \eqref{eq:horizonexpression} for the corresponding values. The plots are for the $\sqrt{s}=200\,\si{\GeV}$ initial conditions.}
    \label{fig:exp4}
\end{figure*}

The late time falloffs of the axial charge can no longer be fitted by $n_5\sim e^{-\Gamma \tau}/\tau$ but the exponent in the denominator and the argument of the exponential are modified. In particular, the axial charge density decays faster than $1/\tau$. Since the chiral magnetic current is reliant on axial charge its late time falloff is also accelerated. We show the falloff in the left plot of figure \ref{fig:exp4}.
\begin{widetext}
In the case of $\Delta=1.25\times 10^{-7}$, the fits to a function of the form $A e^{B\tau+C/\tau}\,\tau^D$ are $$\langle J_\text{CME}\rangle/\epsilon_\infty^{3/4}\sim 0.0026/\tau^{1.35} e^{0.74/\tau-10^{-5}\tau},\, A_v(1)\sim 0.304/\tau^{0.34} e^{0.74/\tau-10^{-5}\tau},\, n_5/\epsilon_\infty^{(3+\Delta)/4}\sim 0.390/\tau^{1.00}\,e^{0.0038/\tau-10^{-5}\tau}\,$$.  
The power law scalings match closely what we expect for the $\Delta=0$. In the case of $\Delta=0.3$, the fits are $$\langle J_\text{CME}\rangle/\epsilon_\infty^{3/4}\sim 0.017/\tau^{2.27} e^{1.88/\tau-0.069\tau}, \,A_v(1)\sim 1.64/\tau^{1.33} e^{2.77/\tau-0.066\tau},\,n_5/\epsilon_\infty^{(3+\Delta)/4}\sim 2.37/\tau^{2.00}\,e^{2.54/\tau-0.068\tau}$$. The power law scalings are seemingly modified. \end{widetext}

One reason for this could be that the charge relaxation rate, which is the factor in the exponent, increases as the magnetic field decreases (for the value of $\alpha$ that we chose in this work; see fig. 1 of \cite{Grieninger:2023wuq}). In particular, in \cite{Grieninger:2023wuq}, we show by considering the QNMs around a static background that $\Gamma\sim c_1-c_2 B^2$. Na\"ively, this leads to a decay of the form $e^{-\Gamma \,\tau}\sim e^{-(c_1 \,\tau-c_2 b^2/\tau)}$ (for $\alpha=6/19$). Moreover, the anomalous dimension increases with $m_s$ ($\Delta)$ but we normalize our quantities to the time independent quantity $\epsilon_\infty$. 

At finite $\Delta$, we were able to establish a horizon formula which correctly reproduces the chiral magnetic current
\begin{equation}\label{eq:horizonexpression}
    \langle J_\text{CME}\rangle=\frac{24\pi^2}{19\,\kappa_5^2}\frac{\alpha}{3(1-\Delta)}A_v(\tau,1)\,B(\tau),
\end{equation}
where the first factor in the product is 1, $A_v(\tau,1)$ is the horizon value of the temporal component of the axial gauge field. Note that since we impose that $A_v(\tau,0)=0$ the horizon value may play the role of an axial chemical potential and a relative factor of 3 also appears in the discussion of covariant and consistent anomaly. The remarkable agreement can be seen in the right side of figure \ref{fig:exp4}.

In figure \ref{fig:exp2}, we show the dependence of axial charge and chiral magnetic current on the collision energy at a small, fixed mass $m_s$. The three different initial conditions outlined in table~\ref{tab:iniexp} correspond to $\sqrt{s}=250\,\si{\GeV}$ (red), $\sqrt{s}=200\,\si{\GeV}$ (black) and $\sqrt{s}=150\,\si{\GeV}$, respectively. Furthermore, we fixed the peaks of the axial charge to the values in the table so that the peak decreases with decreasing the collision energy. The maximum is reached at slightly earlier times for smaller collision energies. Moreover, the chiral magnetic current, which is mostly driven by the magnetic field and axial charge decreases since both decrease with the collision energy. 

\begin{figure*}%[h!]
    \centering
    \subfloat{\includegraphics[scale=0.4]{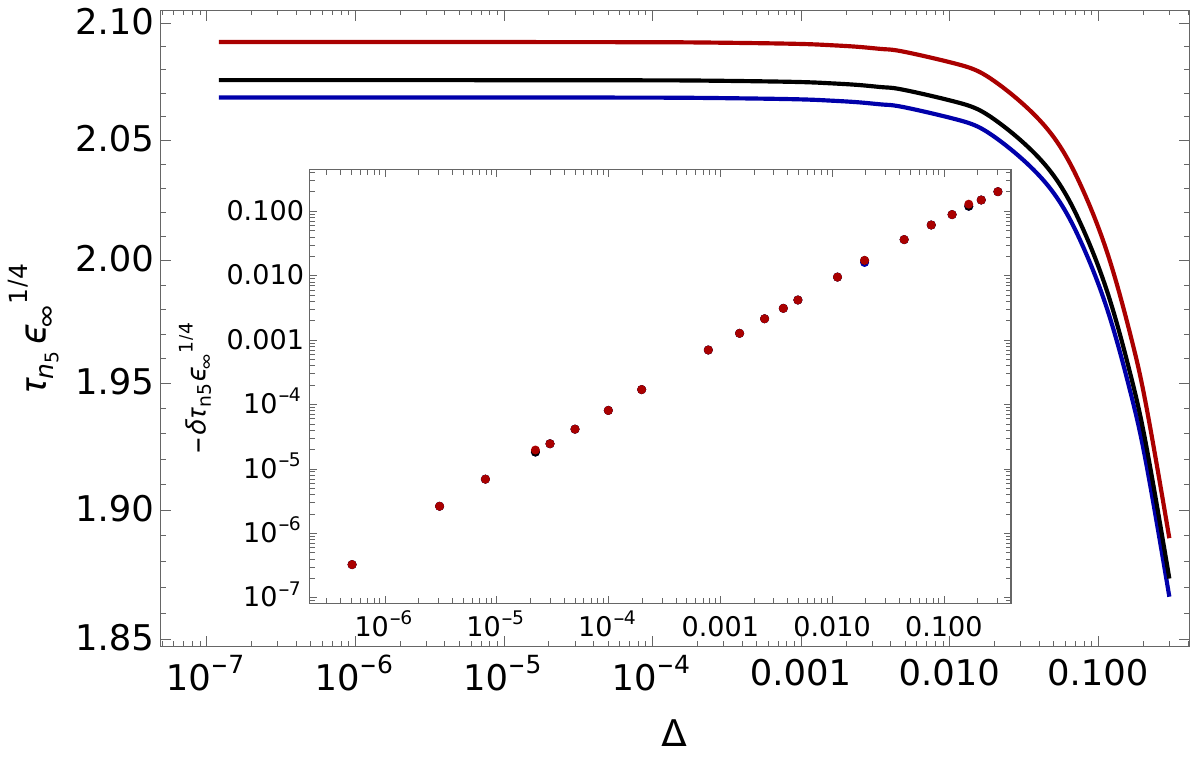}}
    \
    \subfloat{\includegraphics[scale=0.4]{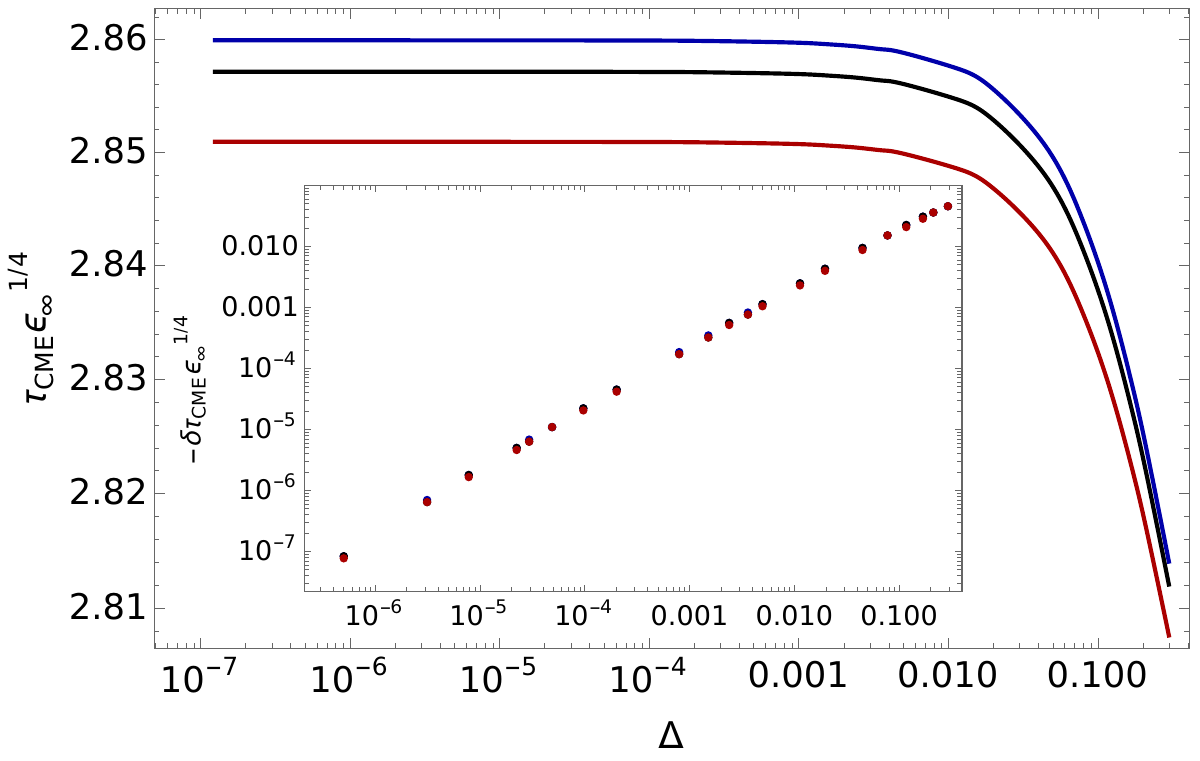}}
    \caption{Build up times (time to the peak) for the axial charge density (left) and chiral magnetic current (right). The inset shows the difference of build up time as a function of $\Delta$ compared to the smallest $\Delta$. The build up times decrease for increasing $\Delta$. The plots are for the $\sqrt{s}=200\,\si{\GeV}$ initial conditions.}
    \label{fig:exp3}
\end{figure*}

The dependence of the maximum on the strength of the non-Abelian anomaly is further detailed in figure \ref{fig:exp3} which shows the time where the axial charge and chiral magnetic current peak as well as the time difference between the peaks as a function of the mass $m_s$. Increasing the mass (or in other words $\Delta$), axial charge and current peak at earlier times. For small enough $m_s$ (or equivalently $\Delta$) the decrease is linear in $\Delta$ as shown by the insets. Moreover, the relative time between the peak of chiral magnetic current and the peak of axial charge increase with $\Delta$ which means that the maximum of the axial charge moves to earlier times faster. We detail the fits in table \ref{tab:inidel}.

\begin{table}
\begin{center}
\begin{tabular}{l| c c c }
\hline &&&\\[-8.5pt]
Beam energy in [GeV]: &$\sqrt{s}=175$&$\sqrt{s}=200$&$\sqrt{s}=250$ 		 \\ \addlinespace\addlinespace
\hline\hline
$\delta\tau^{\text{peak}}_{\text{n}_5}
\epsilon _{\infty }^{1/4}$&$0.836 \Delta^{1.00}$&$0.838 \Delta^{1.00}$&$0.841 \Delta^{1.00}$  \\
$\delta\tau^{\text{peak}}_{\text{CME}} \epsilon _{\infty }^{1/4}$ &$0.218 \Delta^{1.00}$&$0.215 \Delta^{1.00}$&$0.207 \Delta^{1.00}$  \\
$(\delta\tau^{\text{peak}}_{\text{CME}}-\tau^{\delta\text{peak}}_{\text{n}_5})\epsilon _{\infty }^{1/4}$&$0.618 \Delta^{1.00}$&$0.623 \Delta^{1.00}$&$0.634 \Delta^{1.00}$\\
 \addlinespace
\hline 
\end{tabular}
\caption{Change in build up time (time of the peak) with respect to the smallest $\Delta$ as a function of $\Delta$ for $\Delta\ll1$.}
\label{tab:inidel}
\end{center}
\end{table}
\section{Conclusions}\label{sec:conclusions}
We investigated the real-time dynamics of axial charge due to the non-Abelian anomaly and its impact on the chiral magnetic current in detail. 

In the first part of the paper, we focused on a static plasma. The axial charge relaxation rate, which is related to the Chern-Simons diffusion rate (see also\cite{Grieninger:2023wuq}) is determined by the lowest Quasi-Normal Mode (QNM). More precisely it is given by the gap in the imaginary part at zero wave vector. We showed that the axial charge relaxation rate increases for increasing strength of the non-Abelian anomaly $m_s$ and decreases for increasing strength of the Abelian anomaly $\alpha$. Moreover, for $\alpha>0.15$, the axial charge relaxation rate decreases for stronger magnetic fields. Since the axial charge relaxation dynamics is governed by the gap of the QNM at zero wave vector, we can investigate its dynamics in a homogeneous simulation. In explicit time evolution of the static plasma, we first performed a parameter scan. We verified that larger values of $m_s$ lead to faster axial charge relaxation. At large values of $\alpha$ and sufficiently strong magnetic fields, we observe oscillations in the chiral magnetic current and axial charge density. In the case of the axial charge density the oscillations are amplified for increasing $m_s$. Moreover, chiral magnetic current and axial charge density peak faster for increasing $m_s$ (at fixed $B/T^2$ and $\alpha$). If we vary $B/T^2$ at fixed $m_s$ and $\alpha$, then the chiral magnetic current peaks faster for stronger magnetic fields while the axial charge density peaks slower (since the axial charge relaxation rate decreases with increasing the magnetic field). We then considered simulations with initial conditions mimicking the initial conditions of hydrodynamic simulations of RHIC and LHC collisions. For the static plasma, we find that the CME obtained for LHC-like parameters is smaller by a factor of 3 compared to RHIC-like simulations in dimensionless units.

In the second part of the paper, we considered an expanding plasma in which the magnetic field falls off with the inverse proper time (and also the energy density, temperature and pressures are decreasing due to dilution). We performed simulations for different $m_s$ with parameters mimicking collision energies of $\sqrt{s}=250\,\si{\GeV}$, $\sqrt{s}=200\,\si{\GeV}$ and $\sqrt{s}=150\,\si{\GeV}$, respectively. The chiral magnetic current decreases to the lower collision energies due to smaller peak values of the axial charge density.
In the case of an expanding plasma, the Chern-Simons diffusion rate and hence the axial charge relaxation rate are time dependent due to the falloff of the magnetic field. For the value of the Abelian anomaly that we considered, the axial charge relaxation rate increases for decreasing magnetic field and axial charge relaxation is accelerated at later times (since the magnetic field decays as $1/\tau$). The modified relaxation dynamics also impacts the chiral magnetic current which relies on the axial charge and hence is decaying faster. We were able to express the late time decay of the chiral magnetic current in terms of the horizon value of the temporal component of the axial gauge field $A_v(\tau,1)$, the magnetic field, the strength of the Abelian anomaly $\alpha$ and the strength of the non-Abelian anomaly $m_s$. Thus, $A_v(\tau,1)$ mimics the role of an axial chemical potential in our system out of equilibrium with the explicitly broken $U(1)_A$ symmetry. 

We showed that the chiral magnetic current and axial charge density peak faster for increasing $\Delta$ (the decrease in time to the peak is linear in $\Delta$ for small $\Delta$). Since the decrease is faster in case of the axial charge density the difference in peak times of axial charge density and axial current increases. Note that the axial charge density reaches its peak faster and the chiral magnetic current lags.

The first valuable extension of our work would be to consider a finite size system in a box and study interplay of expansion and topological transitions driving the system out of equilibrium. Note that finite volume effects on the CME dynamics were discussed in \cite{Buzzegoli:2023eeo}.

Furthermore, it would be intriguing to explore the time dependence of the energy density on the hydrodynamic side \eqref{eq:hydroeps} considering all potential transport effects arising from strong magnetic fields. This can be achieved by extending the hydrodynamic theory presented in \cite{Ammon:2020rvg} to encompass $U(1)_V\times U(1)_A$. Note that the holographic time evolution on which our results are based already includes all possible transport effects since the energy density evolves according to the bulk equations of motion which capture the full field theory out-of-equilibrium dynamics.

Additionally, investigating the impact of dynamical (Abelian) magnetic fields along the lines of \cite{Grozdanov:2017kyl,Ahn:2022azl,Baggioli:2023oxa} would be of great interest in the light of recent developments in magnetohydrodynamics \cite{Das:2022fho,Das:2022auy,Landry:2022nog}.

In the context of realistic heavy-ion collisions, where the plasma undergoes rapid expansion, we have focused on studying the homogeneous dynamics of the topological axial charge. To enhance our understanding, it would be highly valuable to expand this analysis to include spatial dynamics as for example presented in \cite{Grieninger:2023wuq} which would allow making the magnetic field time dependent.

In view of the beam energy, it would be valuable to perform an energy scan in holography focusing on low temperatures. However, our current holographic model is limited to sufficiently high temperatures in the plasma phase (around $T\sim300\si{\MeV}$), as we have not incorporated a realistic behavior of the entropy density close to the QCD phase transition. To explore dynamics at lower temperatures, it may be beneficial to employ more sophisticated holographic models for QCD such as V-QCD (see \cite{Jarvinen:2022doa} for a recent review).

In future publications, we intend to address some of these intriguing questions.

\section*{Acknowledgments}
We thank Casey Cartwright, Dima Kharzeev, Karl Landsteiner, Shuzhe Shi and Ismail Zahed for helpful discussions.
We are especially grateful for the insightful discussions with Shuzhe Shi regarding initial conditions in hydro simulations, and we also appreciate the valuable input from Casey Cartwright in our discussions about \cite{Cartwright:2021maz}.

This work was supported by the Office of Science, Office of Nuclear Physics, U.S. Department of Energy under Contract No. DE-FG88ER41450. S.M.T. was supported through the Grants No. CEX2020-001007-S and PGC2018-095976-B-C21 funded by MCIN/AEI/10.13039/501100011033 and by ERDF ``A way
of making Europe,'' as well as by an FPI-UAM predoctoral fellowship.

\appendix
\section{Small charge approximation}\label{app:smallcharge}
\begin{figure*}
    \centering
  \includegraphics[width=0.49\linewidth]{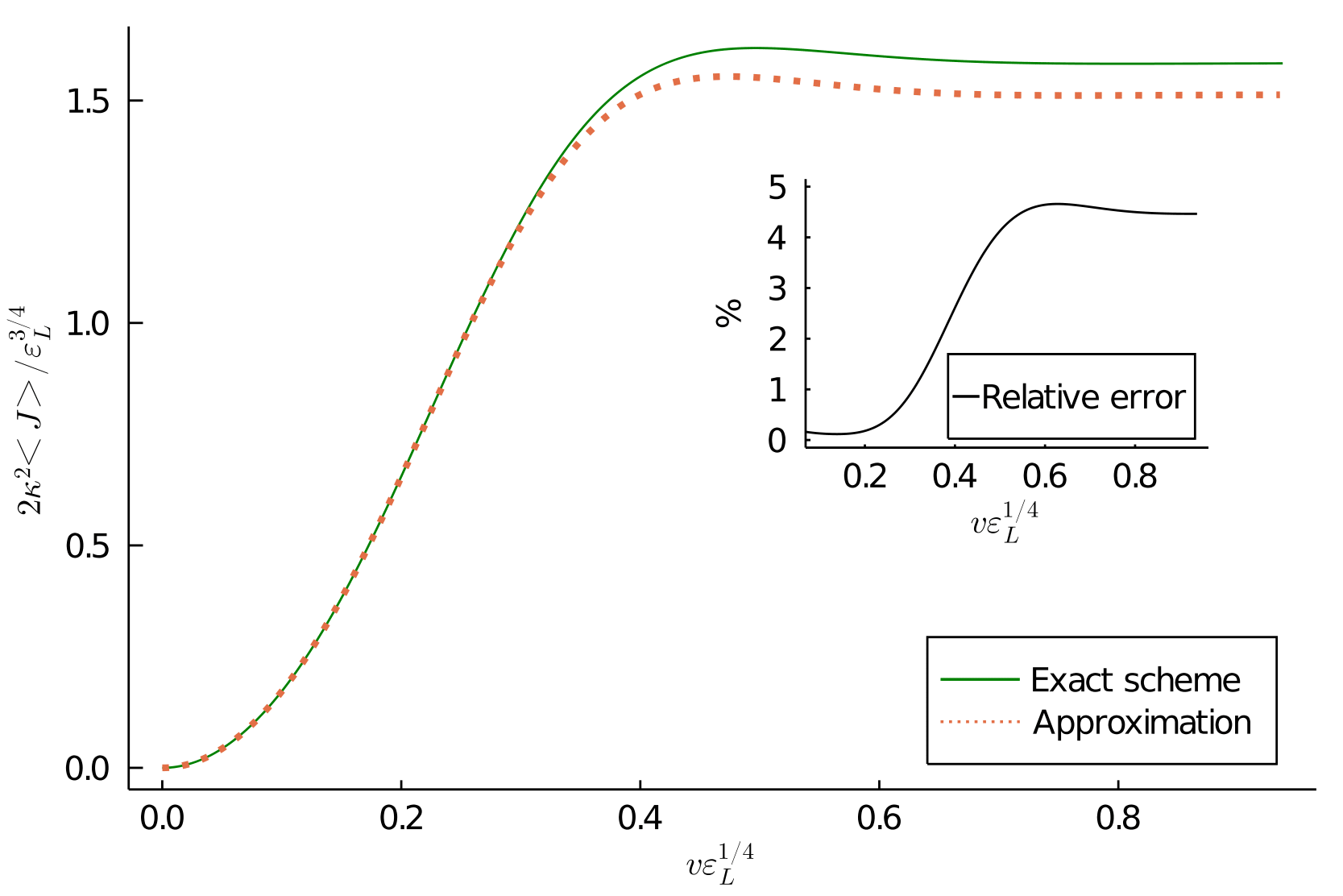}
    \,
   \includegraphics[width=0.49\linewidth]{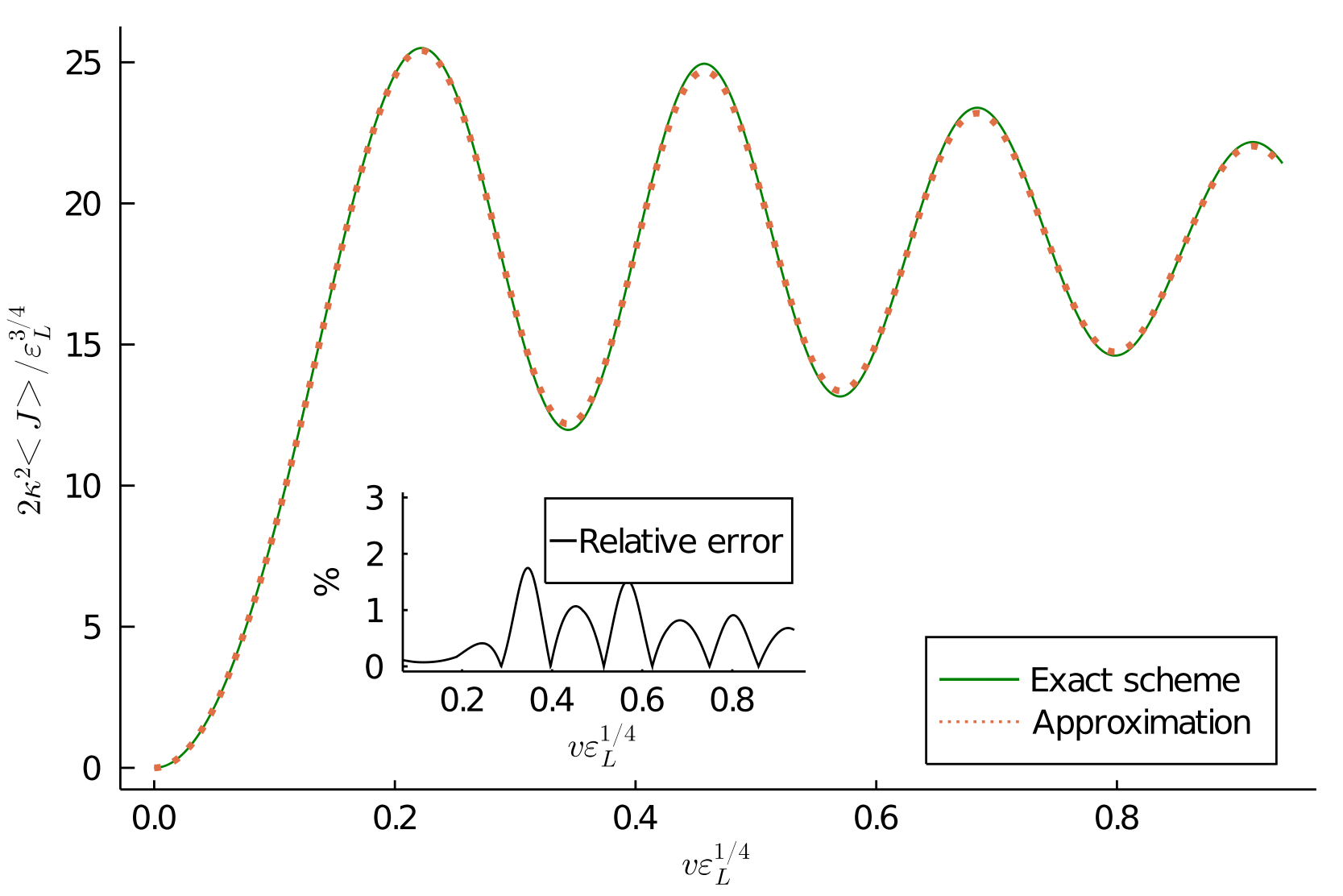}
    \caption{Comparison of vector current $q=1$ and relative error for $B=0.02$ (Left) and $B=1$ (Right).}
    \label{fig:1}
\end{figure*}

We have claimed that this approximation is to be trusted only when charge is small. To be precise, it has to be small compared to the other scales in the problem, e.g. temperature, magnetic field, etc. It suffices that it is small compared to one of them so that we can expand around the dimensionless ratio of charge and that particular scale. Let us discuss the dimensionless ratio $n_5/T^{3+\Delta}$. In holography, temperature appears in the combination $ 2\pi T$, so it seems reasonable to guess that the approximation is valid so long as $\dfrac{n_5}{( 2\pi T)^{3+\Delta}}\lesssim 0.5\,$. We proceed now to verify this intuition numerically. The full backreacted system has been solved for $\Delta=0$ in \cite{Ghosh:2021naw} so we may compare an exact result with our approximation. Two examples are shown in figures \ref{fig:1}, corresponding to $n_5/( 2\pi T)^3 = \{0.15\, , 0.14\} \,$ and $B/T^2 = \{ 0.228 \,, 10.76 \} \, $ respectively. In both cases the relative error is below $5\%$, confirming our expectations. We will later see that the parameters chosen lie in the small charge regime.

\section{QNMs and late time behavior}\label{app:qnmslatetime}
\begin{figure*}
    \centering
    \subfloat{\includegraphics[scale=0.35]{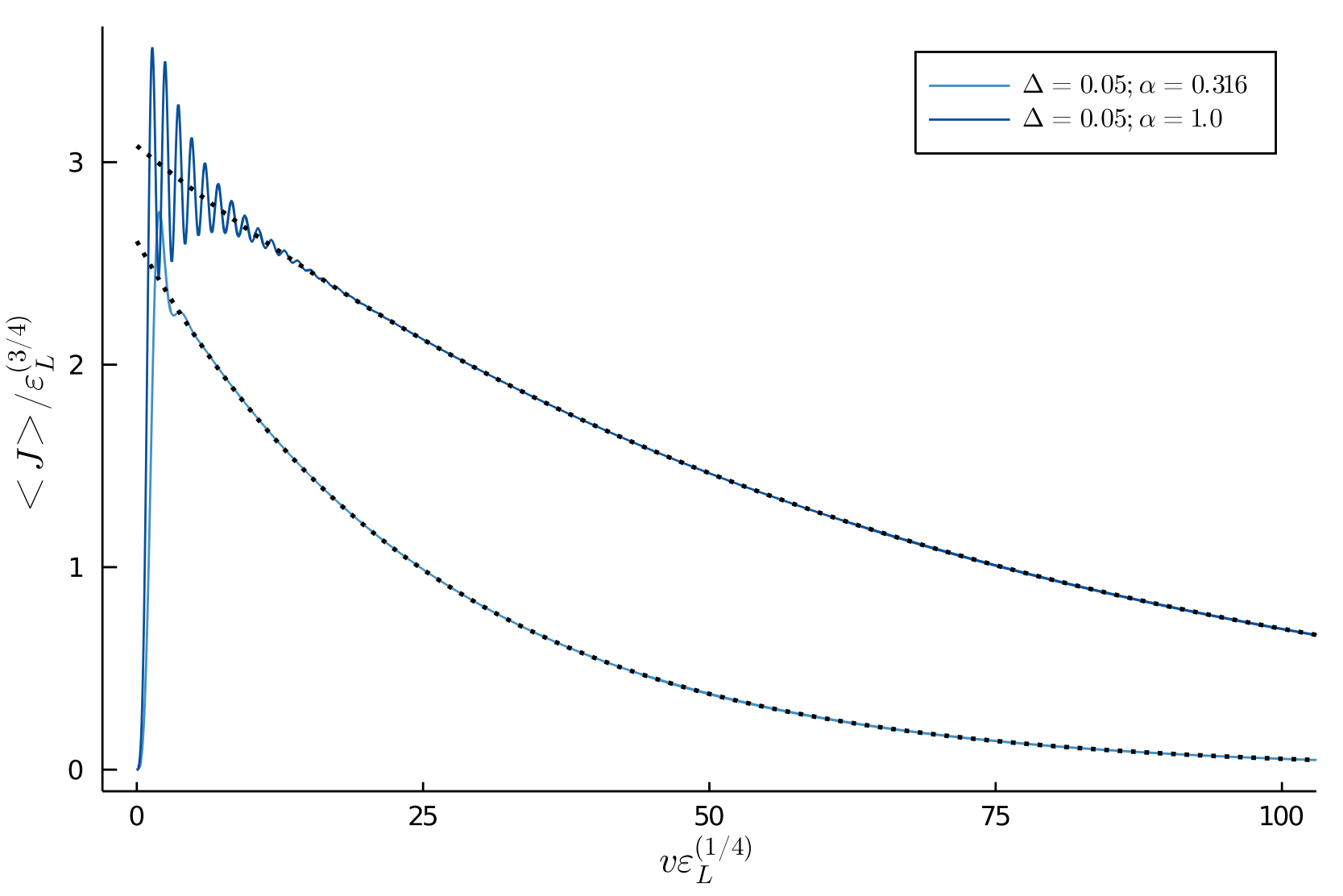}}
    \qquad
    \subfloat{\includegraphics[scale=0.35]{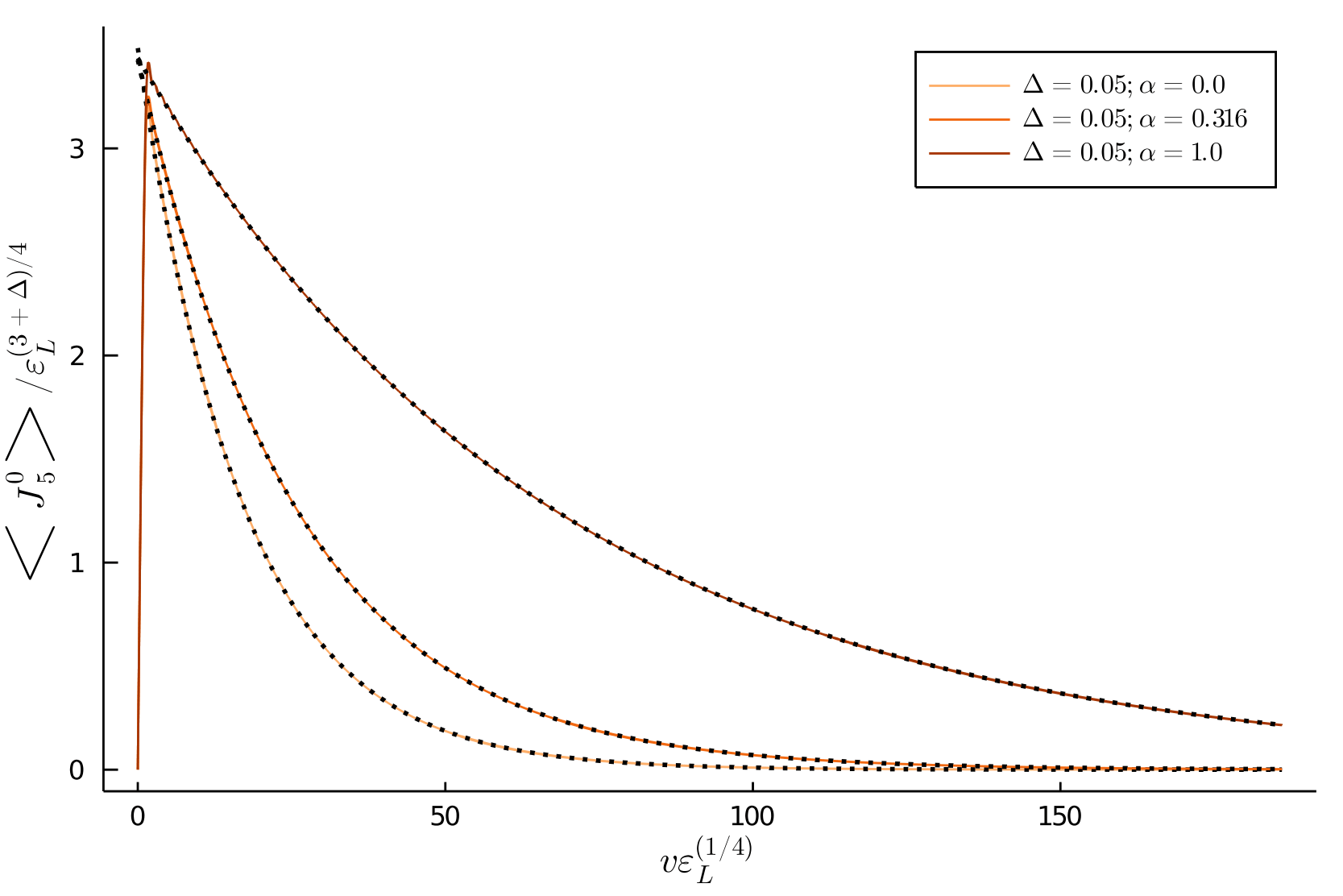}}
    \caption{(Left) Chiral magnetic effect. (Right) Axial charge. (Both) Dashed lines correspond to the lowest quasinormal mode. % for the given state computed in \eqref{eq:qnm}. 
    Simulations are for fixed $B/T^2 = 10.09$ and $\Delta=0.05$. In the initial state, we impose $J_{CME}^{ini}=n_5^{ini}=0$ and $\Dot{q}_5^{ini}=1$.}
    \label{fig:qnm1f}
\end{figure*}

In this appendix, we explicitly show that the chiral magnetic current and axial charge density falloff exponentially at late times in the static plasma. Furthermore, by overlaying the falloff with the lowest QNM we verify that the exponential falloff is indeed governed by the lowest QNM which describes the axial charge relaxation. An example of this is depicted in figure \ref{fig:qnm1f}. The figures shows that the chiral magnetic current and the axial charge density decay exponentially in time (in the static plasma) due to the explicit breaking of the axial $U(1)$.
\section{Additional figures pertaining to the static plasma simulations}\label{app:parameterscanstatic}

\begin{figure*}
	\centering
	\subfloat{\includegraphics[scale=0.37]{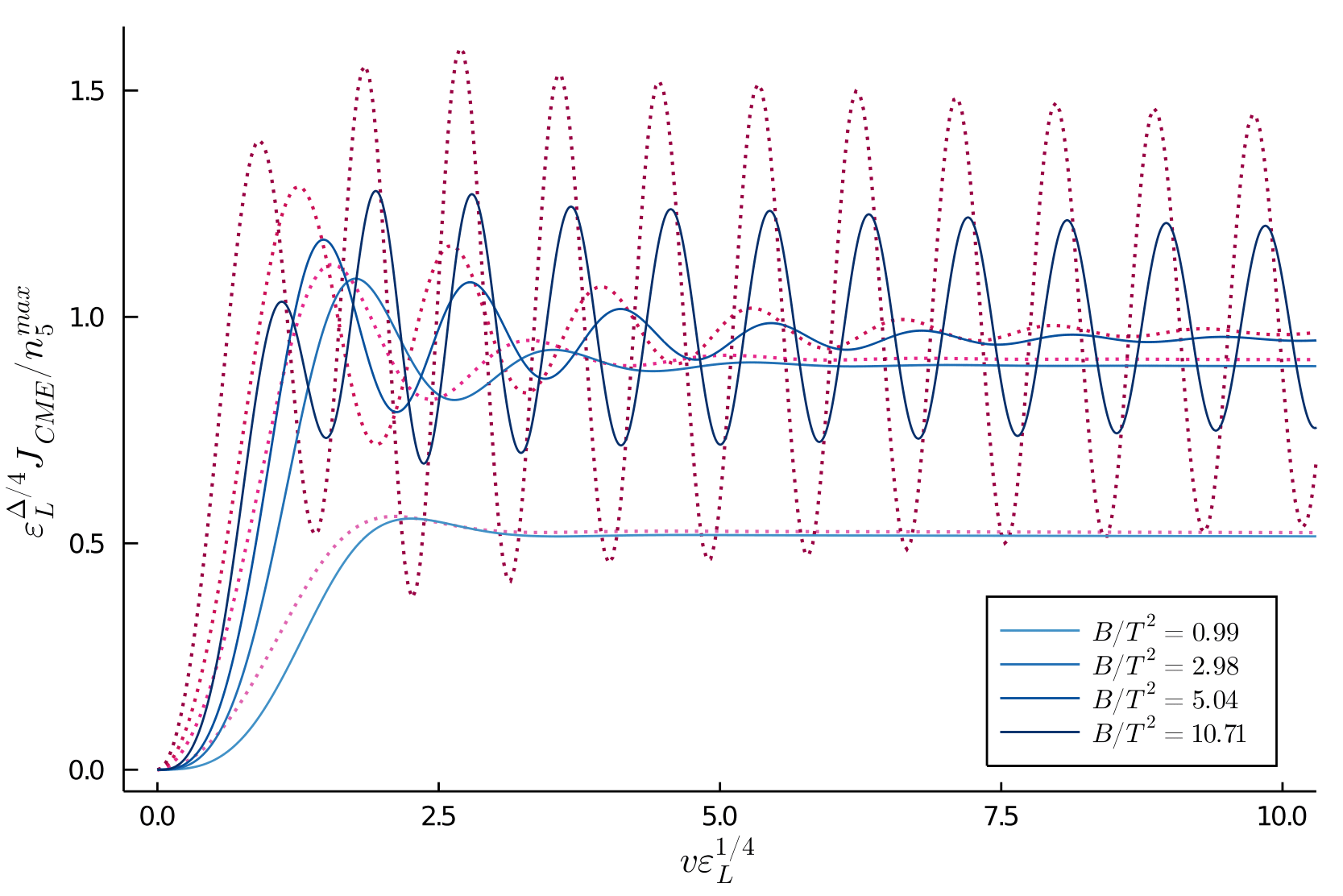}}
	\qquad\vspace{0.4cm}
	\subfloat{\includegraphics[scale=0.37]{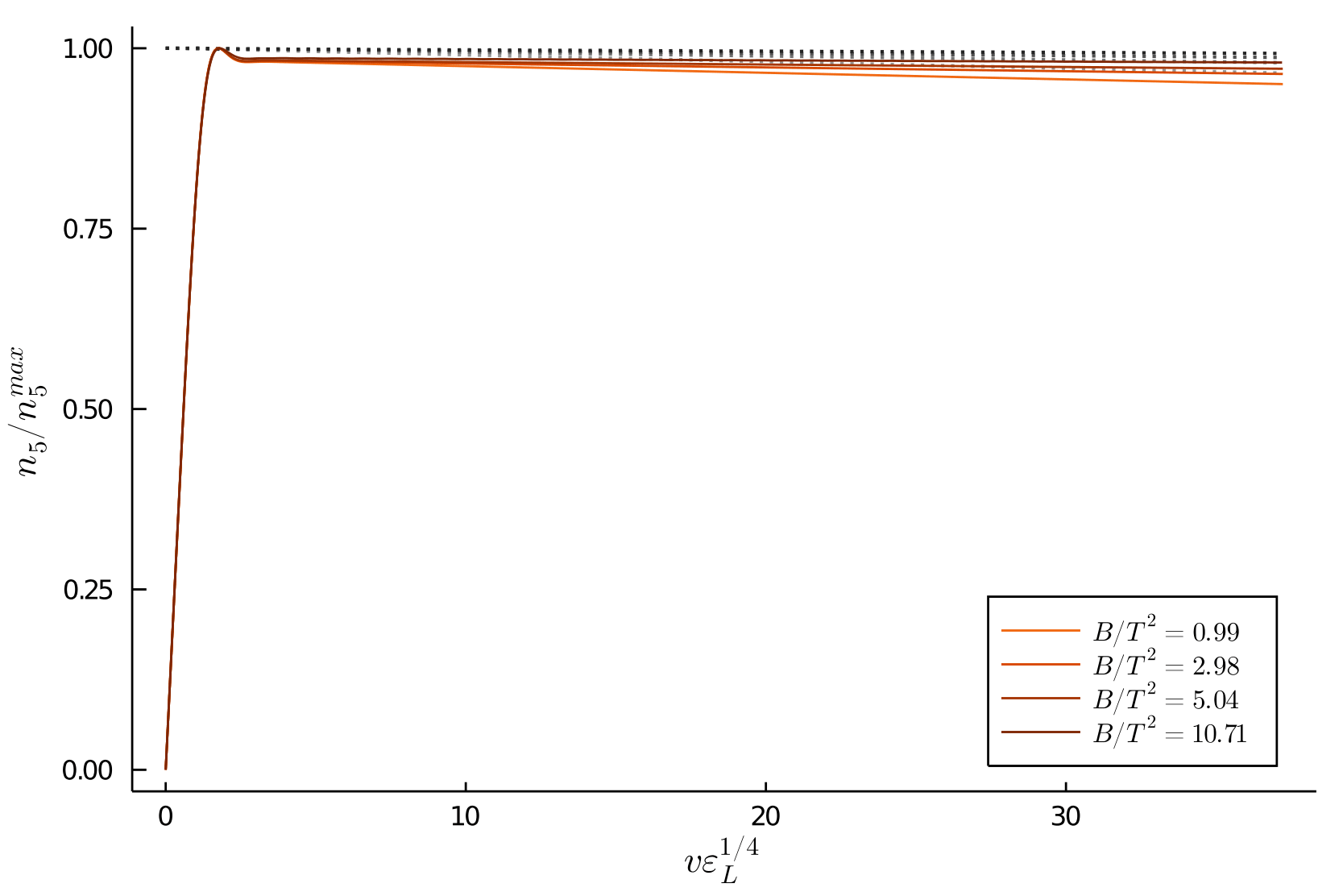}}
	\qquad
	\subfloat{\includegraphics[scale=0.37]{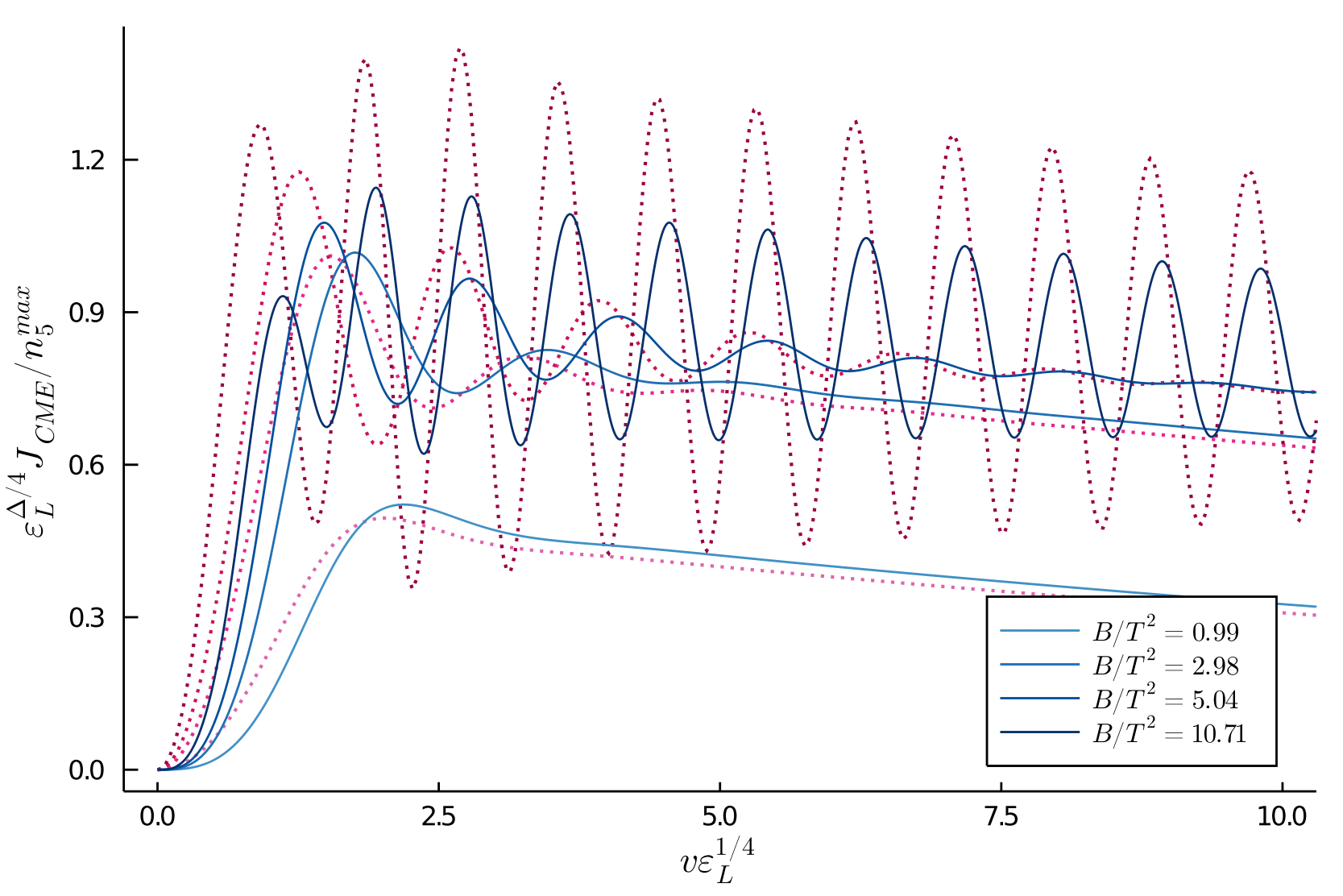}}
	\qquad\vspace{0.4cm}
	\subfloat{\includegraphics[scale=0.37]{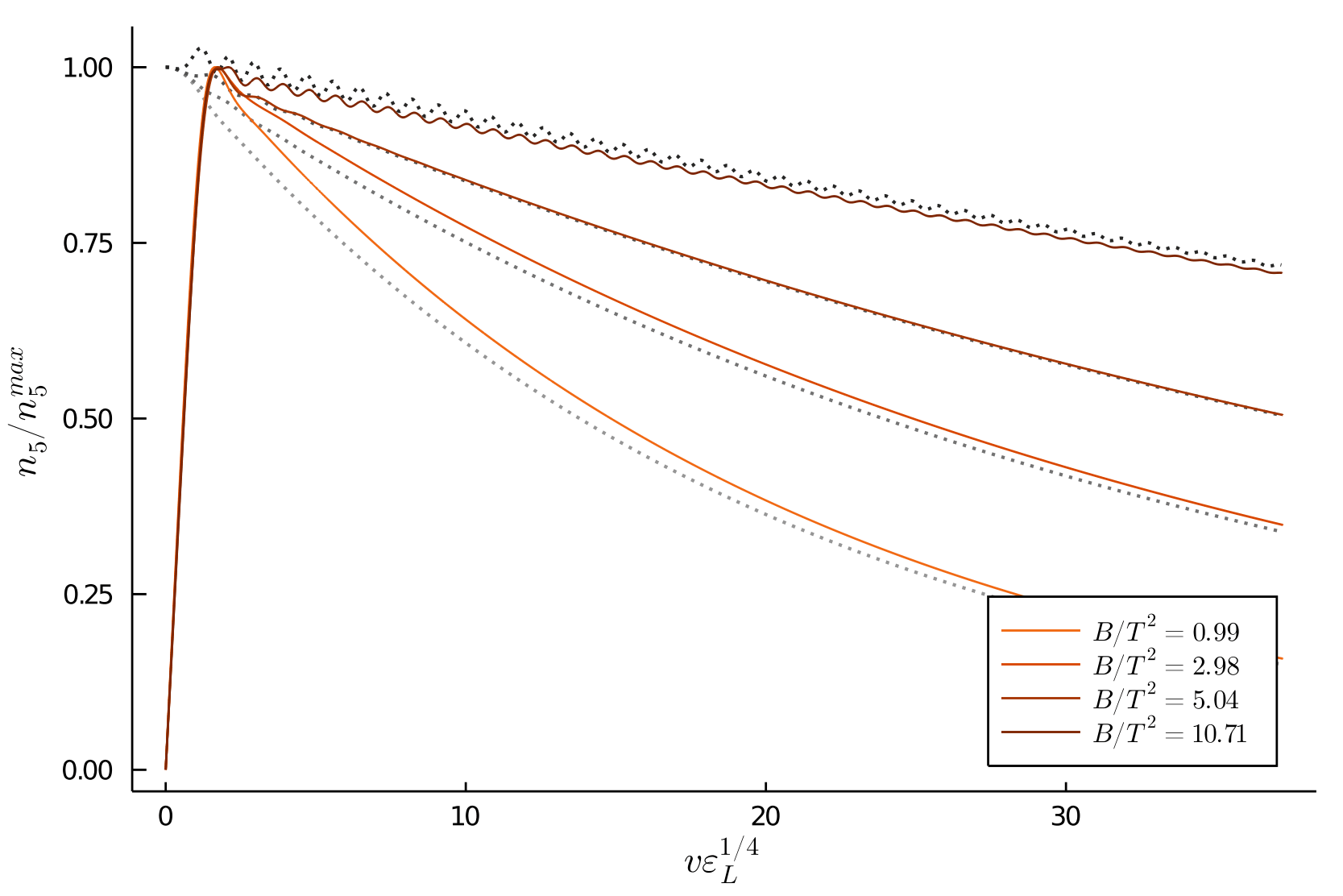}}
	\caption{(Left) Chiral magnetic effect and (Right) axial charge for simulations with $\alpha=1.5$ for $\Delta=0.001$ (top) and $\Delta=0.05$ (bottom), respectively. Solid lines correspond to initial state (A) whereas dashed lines correspond to state (B). Solid lines are normalized with respect to the maximum value of $n_5$, whereas dashed lines are normalized to the initial value of $n_5$.}
	\label{fig:adim15}
\end{figure*}

\begin{figure*}[h!]
	\centering
	\subfloat{\includegraphics[scale=0.37]{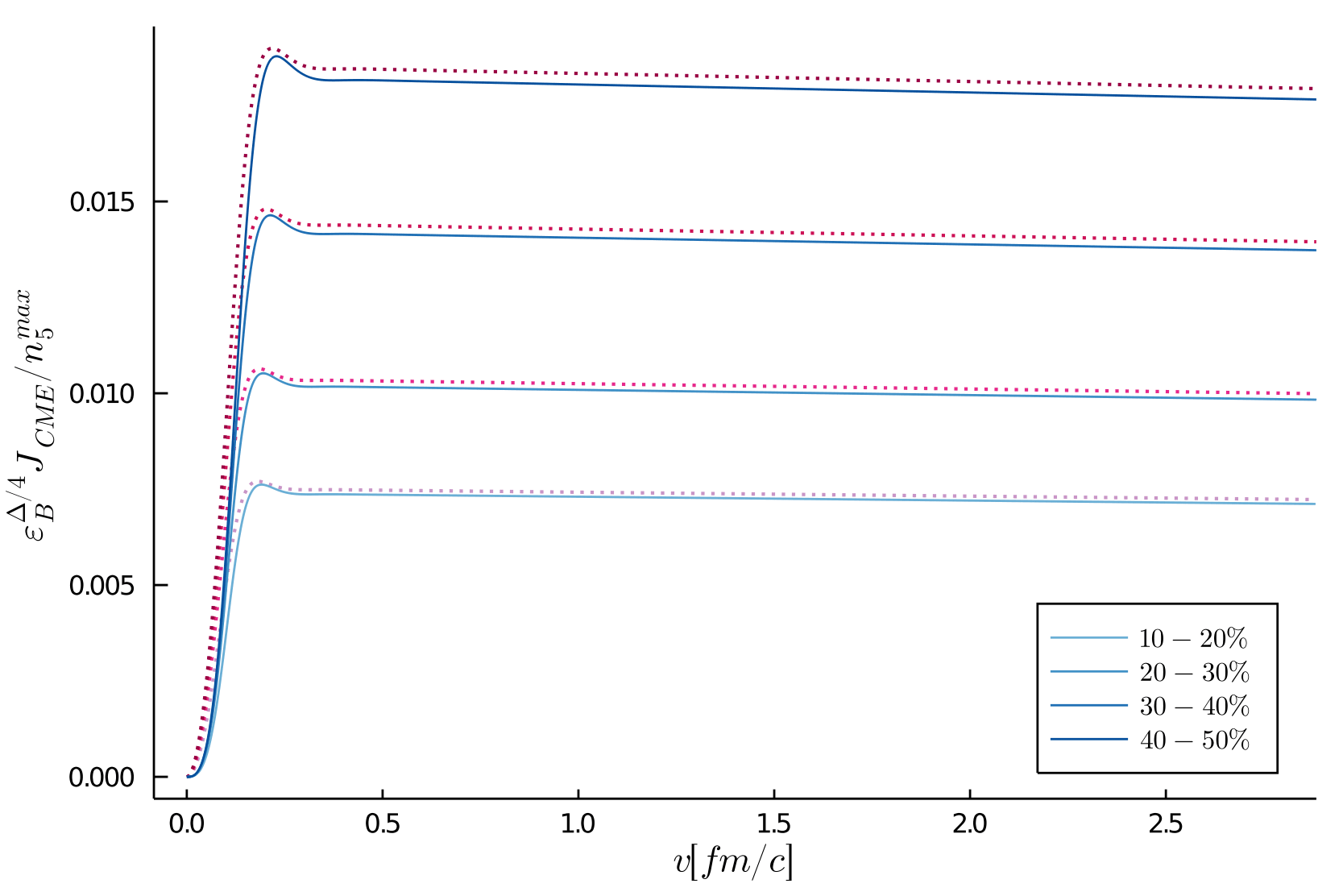}}
	\qquad\vspace{0.4cm}
	\subfloat{\includegraphics[scale=0.37]{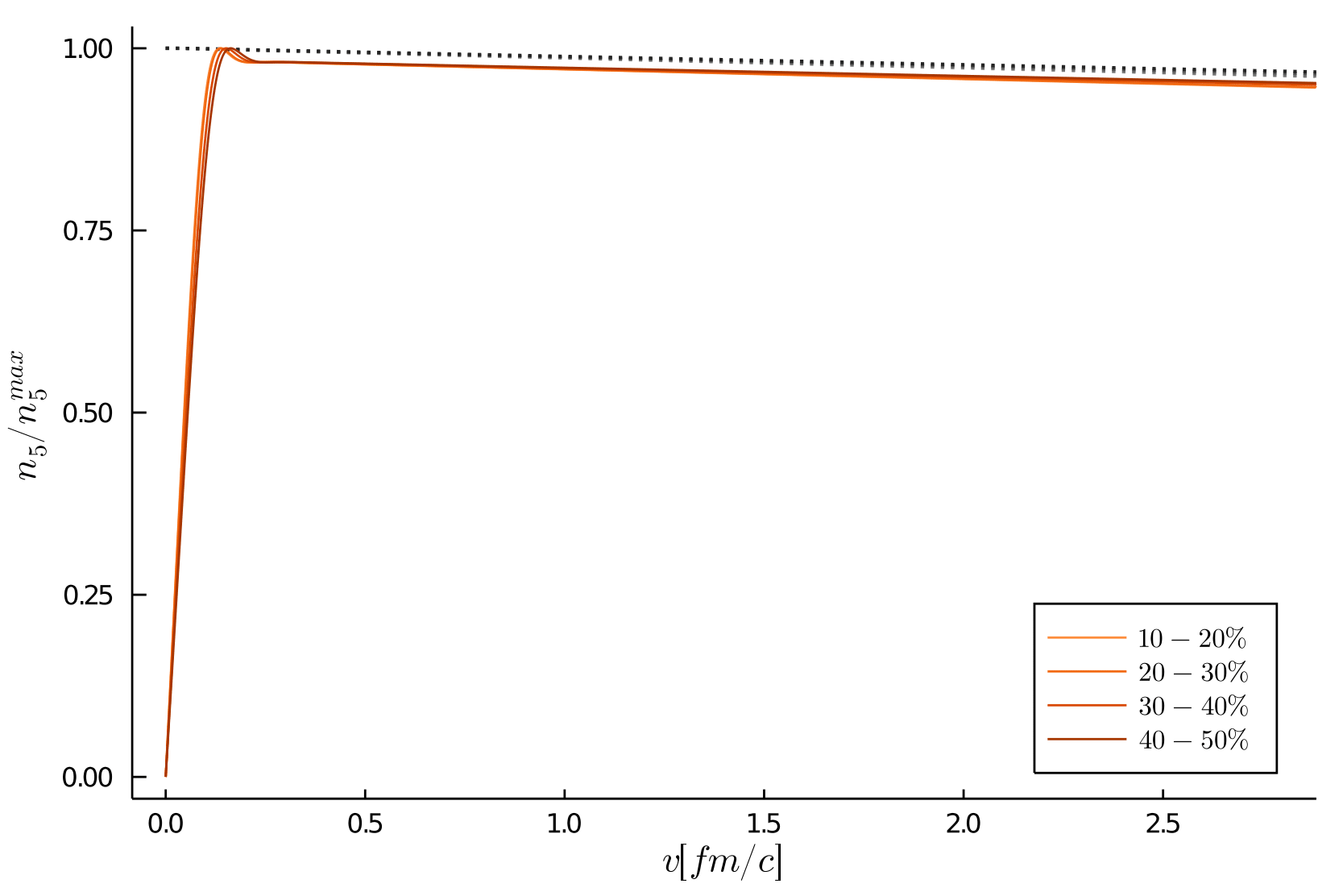}}
	\qquad
	\subfloat{\includegraphics[scale=0.37]{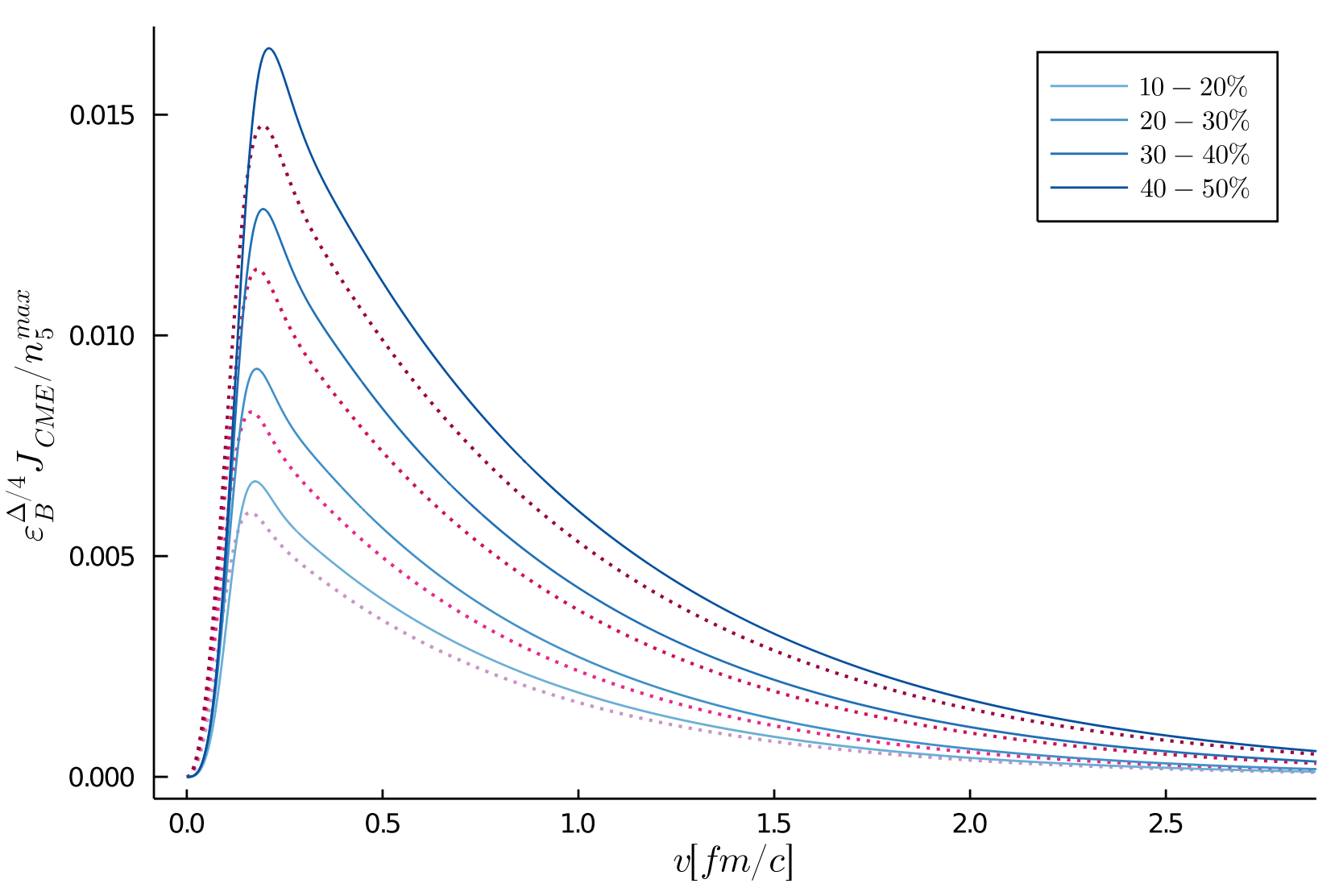}}
	\qquad\vspace{0.4cm}
	\subfloat{\includegraphics[scale=0.37]{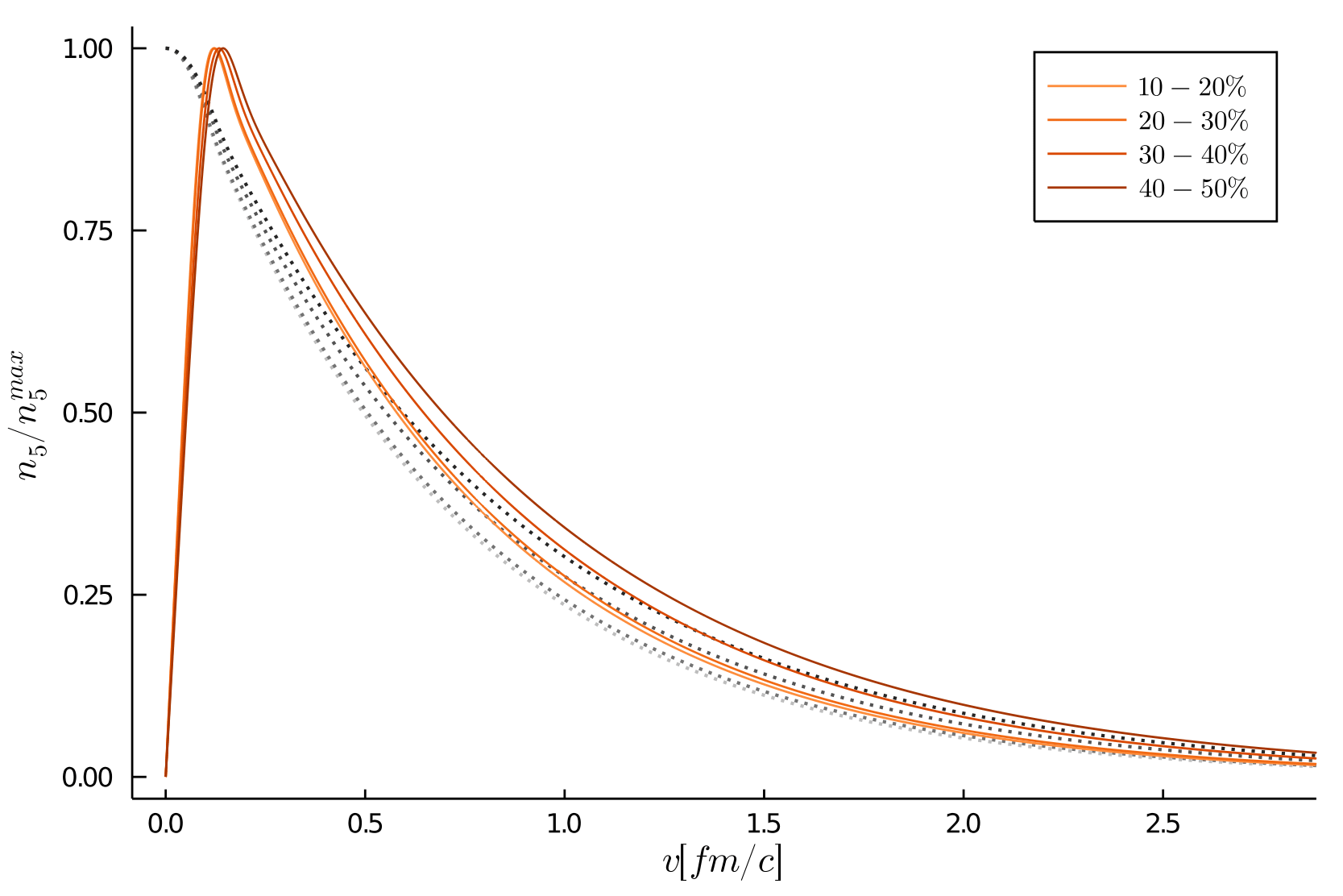}}
	\caption{(Left)  Chiral magnetic effect and (Right) axial charge for simulations with LHC-like parameters as a function of centrality. We set $\alpha=\frac{6}{19}$ as well as $\Delta=0.001$ (top) and $\Delta=0.11$, respectively. Solid lines correspond to initial state (B) whereas dashed lines correspond to state (A). The labeling refers to data in table \ref{tab:2} and it applies to both solid and dashed lines in a correlated manner. }
	\label{fig:LHCn}
\end{figure*}

The results for simulations with $\alpha=1.5$ are shown in figure \ref{fig:adim15}. The dashed and solid lines correspond to initial states (A) or (B), respectively. Each plot is done at a fixed value of $\Delta$, which increases from top to bottom, and within each plot we study the magnetic field dependence. We show the evolution in dimensionless time, i.e. time normalized to energy density $v\epsilon_L^{1/4}$. Clearly, higher values of $\Delta$ result into faster dissipation, in agreement with the fact that $\Delta$ measures the degree of nonconservation of the axial current. On top of that, dissipation is more significant for lower magnetic fields. Besides, both observables (vector current and axial charge) display oscillatory behavior, which is more prominent as the magnetic field is increased. These features are well described by the quasinormal modes computed in \ref{sec:qnm}. The presence/absence of oscillations seems to be independent of the value of $\Delta$ for the vector current. On the contrary, oscillations in axial charge become more important as we increase the value of $\Delta$. 

In addition to that, we observe two interesting phenomena concerning the initial time response. The first of them the initial time response of the axial charge seems to be insensitive to the magnetic field, that is all curves overlap initially. This is linked to the fact that during an arguably short time, the behavior of the axial charge is solely dictated by the initial state given by hand. In particular, for the initial state (A), the axial charge develops a \textit{plateau} before dissipation kicks in, whereas initial state (B) it follows a linear behavior. The duration of either of them is roughly the same: $v\epsilon_L^{1/4}\sim0.5$ This is clearly an artifact of the initial state, for instance in (A) we are setting $A_t(0,u) = n_5(0) u^{2+\Delta}$, which is tantamount to demanding that the time derivatives of $n_5$ vanish, as these enter the asymptotic expansion at higher orders in $u$. This shows that the evolution of axial charge is initially strongly dependent on the assumptions made. However, the qualitative features discussed are expected to be valid for generic out of equilibrium\footnote{At equilibrium the solution is trivially zero and there are no features to be discussed.} initial states which lead to sufficiently small $n_5^\text{max}$, approximately $n_5\leq 0.5\,(\pi T)^{3+\Delta}$. The second phenomenon we would like to highlight is precisely the time response of the vector current: it builds up faster as we increase the magnetic field. The same behavior was found in \cite{Ghosh:2021naw}. Roughly, the explanation is that at high magnetic fields, the gap between the lowest and first Landau levels increases and the fermions remain in the lowest Landau level. Then the physics becomes effectively $1+1$ dimensional and there is an operator relation between the vector current and axial charge, implying that the response should be instantaneous. We refer the reader to \cite{Ghosh:2021naw} for a more detailed discussion. 

Both initial states are qualitatively similar. The most remarkable differences are that in (A) the vector current reacts faster than in (B) for the same magnetic field, and that in (A) the amplitude of the oscillations is significantly bigger. The first difference is explained by the fact that the vector current takes some time to react to changes in the axial charge, thus in (B) axial charge has to build up and then the vector current responds, whereas in (A) there is some initial charge and the vector current may develop accordingly.

A final comment on the initial state dependence/independence concerns ``sharp'' initial states. As shown in \cite{Fuini:2015hba}, even for "sharp" initial states, the effect of the nonlinearities (which we are neglecting here by linearizing the equations) is expected to be small, thus not modifying our discussion. The response of the chiral magnetic current also becomes less dependent on the initial state as we decrease the value of the Chern-Simons coupling $\alpha$.

\bibliography{main}

\end{document}